\documentclass{statsoc}

\usepackage[a4paper]{geometry}
\usepackage{graphicx,graphics,bm,epstopdf}
\usepackage[textwidth=8em,textsize=small]{todonotes}
\usepackage{amsmath,amsfonts,amssymb,hyperref,verbatim}
\usepackage{natbib}
\usepackage[para,online,flushleft]{threeparttable}
\usepackage{dsfont}
\usepackage{algorithm}
\usepackage{algpseudocode}
\usepackage{multirow}
\usepackage{longtable}
\usepackage{booktabs}
\usepackage{pdflscape}
\usepackage{rotating}
\usepackage{mathtools}

\usepackage{etoolbox}
\makeatletter
\patchcmd{\@makecaption}
  {\parbox}
  {\advance\@tempdima-\fontdimen2} 
  {}{}
\makeatother

\renewcommand{\thesection}{\arabic{section}}

\title[Hematopoiesis network inference from RT-qPCR data]{Sparse inference of the human hematopoietic system from heterogeneous and partially observed genomic data}

\author[Sottile G. {\it et al.}]{Gianluca Sottile}
\address{University of Palermo, Department Economics, Business and Statistics, Palermo, Italy.}
\email{gianluca.sottile@unipa.it}
\author[Sottile G. {\it et al.}]{Luigi Augugliaro}
\address{University of Palermo, Department Economics, Business and Statistics, Palermo, Italy.}
\author[Sottile G. {\it et al.}]{Veronica Vinciotti}
\address{University of Trento, Department of Mathematics, Trento, Italy}
\author[Sottile G. {\it et al.}]{Walter Arancio}
\address{Advanced Data Analysis Group, Fondazione Ri.MED, Palermo, Italy}
\author[Sottile G. {\it et al.}]{Claudia Coronnello}
\address{Advanced Data Analysis Group, Fondazione Ri.MED, Palermo, Italy}

\begin{document}

\begin{abstract}
Hematopoiesis is the process of blood cell formation, through which progenitor stem cells differentiate into mature forms, such as white and red blood cells or mature platelets. While the precursors of the mature forms share many regulatory pathways involving common cellular nuclear factors, specific networks of regulation shape their fate towards one lineage or another. In this study, we aim to analyse the complex regulatory network that drives the formation of mature red blood cells and platelets from their common precursor. To this aim, we develop a dedicated graphical model which we infer from the latest RT-qPCR genomic data. The model also accounts for the effect of external genomic data. A computationally efficient Expectation-Maximization algorithm allows regularised network inference from the high-dimensional and often only partially observed RT-qPCR data. A careful combination of alternating direction method of multipliers algorithms allows achieving sparsity in the individual lineage networks and a high sharing between these networks, together with the detection of the associations between the membrane-bound receptors and the nuclear factors. The approach will be implemented in the \texttt{R} package \texttt{cglasso} and can be used in similar applications where network inference is conducted from high-dimensional, heterogeneous and partially observed data.
\end{abstract}


\section{INTRODUCTION}

Human hematopoiesis is a complex cellular process that occurs continuously during the human lifetime and that produces cellular blood components from hematopoietic stem cells. In dedicated niches in the bone marrow, a mixed population of precursors at different levels of differentiation coexists with a functional stroma and it physiologically orchestrates, via cross-regulation, the maintenance of stem-like immature precursors and the commitment towards specific cell lineages, together with the differentiation, the maturation and the release of the mature forms. Recent studies have focussed on the megakaryocyte-erythroid progenitor (MEP) \citep{psaila2016}. The MEP starts from an early immature precursor (pre-MEP) and then commits itself towards either an erythroid lineage (E-MEP), that will become red blood cells, or a megakaryocyte lineage (MK-MEP), that will produce mature platelets. While these precursors have many  differentiation pathways in common, which are driven by key proteins called nuclear factors, specific networks of regulation shape their fate towards one lineage or another. The precise disentanglement of the pathways governing the commitment of the two populations is a difficult task and has received recent attention~\citep{Dore:2011us}.  

While nuclear factors are key players in the late stages of hematopoiesis, they do not act in isolation. Many studies have shown how membrane-bound receptors are also important in the identification and classification of hematopoietic cells~\citep{Cheng:2020uq}. Indeed, membrane-bound receptors play a pivotal role in the relationship with the surrounding cellular microenvironment, and their activation is transduced to the nucleus to reprogram or fine tune the cell fate. As well as inferring the network of regulation among nuclear factors, the proposed method aims to quantify the effect of these membrane-bound receptors on the nuclear factor activities. From a modelling point of view, this creates a rather complex picture with a number of challenging features: firstly, data at the level of nuclear activities (response variables) and at the level of membrane receptor activities (predictor variables or generally external covariates), with potentially high dimensionality on both; secondly, dependence structures both between and within the two sets of variables; thirdly, the presence of multiple biological conditions (Pre-MEP, E-MEP and MK-MEP) which have many features in common but also some key differences. In addition, the reverse transcription quantitative real-time PCR (RT-qPCR) data that will be used for the study, generated by  \cite{psaila2016}, is characterized by a large percentage of missing data, which is typical for this type of data. This will provide further computational challenges, that will be met with the use of a suitably developed Expectation-Maximization (EM) algorithm ~\citep{DempsterEtAl_JRSSB_77}.

Motivated by the applied setting of blood cell formation, we develop a graphical modelling approach that accounts for lineage-dependent regulatory networks, both at the level of the response variables (nuclear activities) and of the covariates (membrane receptor activities) as well as lineage-dependent associations between the two. An added layer of complexity comes from the fact that the data are only partially observed, a feature that is typical of RT-qPCR data but that is common in transcriptomic studies in general, particularly when measurements are made at different levels or under different biological conditions. Penalised estimators developed for the standard and conditional Gaussian graphical model setting under missingness~\citep{AugugliaroEtAl_BioStat_20,AugugliaroEtAl_CompStat_20,StadlerEtAl_StatComp_12} will be extended to the case of multiple conditional Gaussian graphical models, with the development of an EM algorithm that accounts for missingness both at the response and covariate levels. Imposing sparsity within the two levels as well between the two levels provides a challenging computational setting: we propose a careful combination of regularized inferential procedures at the different levels, which we solve based on suitably developed and efficient alternating direction method of multipliers (ADMM) algorithms \citep{boyd2011distributed}. On one side, the methodology aims to identify a common shared network among all the differentiation states; on the other side, it aims to elucidate specific features that characterize the most undifferentiated state (Pre-MEP) and each of the committed lineages (E-MEP and MK-MEP). 

\subsection{Related works}

Inference of the dependence structure of random variables, such as the nuclear activities in the applied setting just described, falls within the remit of graphical models. Among these, Gaussian graphical models (GGMs) are commonly used for multivariate continuous random variables, thanks to their ability to depict  highly complex conditional dependence structures among the random variables in a parsimonious way. Among the penalized estimators proposed in the literature, the  graphical lasso (glasso)~\citep{YuanEtAl_BioK_07} is undoubtedly the most famous one. Recently, the usage of complex sampling schemes and experimental settings have motivated various extensions to account for the heterogeneity in the observed data. 

A first strand of extensions has looked at the case of data measured from different sources or under different conditions, such as the undifferentiated cellular state and the two committed lineages in the formation of blood cells described above. In these cases, a comprehensive description of the data generating process can be obtained by fitting a collection of GGMs, otherwise called a \textit{multiple GGM}, with one network associated to each condition but with the different networks sharing some common structures, such as the presence or intensity of the dependencies.  To accomplish this joint estimation problem, a number of authors have developed specific penalized approaches \citep{GuoEtAl_BioK_11,DanaherEtAl_JRSSB_14,ZhuEtAl_JASA_14,LeeEtAl_JMLR_15}. Of particular notice for the method proposed in this paper is the double-penalized estimator, which we refer to as joint glasso, introduced by~\cite{DanaherEtAl_JRSSB_14}.

The joint glasso approach focusses mainly on network estimation and treats the expected values as nuisance parameters. A second strand of extensions has concentrated on modelling the expected values, by capturing dependencies with external data at the mean level. Penalized methods developed to analyze this kind of heterogeneous data are built on the notion of \textit{conditional GMMs}~\citep{LaffertyEtAl_ICML_01}, which, in turn, are based on the assumption that the covariates affect the multivariate density only through a linear model on the expected value. The penalized estimators that have been proposed to explore the sparse structure of a conditional GGM typically use two specific penalty functions: one that induces sparsity in the regression coefficient matrix, capturing the association between the response variables and the external covariates, and a second one that encourages sparsity in the dependence structure between the response variables \citep{RothmanEtAl_JCGS_10,YinEtAl_AOAS_11,LiEtAl_JASA_12,YinEtAl_JMA_13,Wang_StatSinica_15,ChenEtAl_JASA_16,ChiquetEtAl_StatComp_17}. These methods normally assume that the network describing the dependencies between the response variables is independent of the covariates. A small number of studies have proposed further extensions of these estimators to the case of \textit{multiple conditional GGMs}, essentially providing a bridge with the multiple GGM methods described above \citep{ChunEtAl_Frontiers_13,HuangEtAl_IEEE_TNNLS_18}. In all these methods, the covariates are not treated as random variables and their dependence structure is not explored.

\subsection{Overview of the paper}

In Section~\ref{sec:data_description} we present the motivating example and provide some summary statistics. Section~\ref{sec:model} describes the details of the proposed Gaussian graphical model. Section~\ref{sec:inference} derives the likelihood under the case of partially observed data, both at the response and covariate levels, and presents the penalised likelihood that is used as objective function. Section \ref{sec:algorithm} details the steps of the EM algorithm and of the ADMM algorithms that make up the different sub-routines. Section \ref{sec:simulation} describes an extensive simulation study where we evaluate the performance of the method in terms of network recovery and parameter estimation, and compare it with existing methods. In Section~\ref{sec:realdata} we provide an application of the methodology to the inference of the hematopoietic system and, finally,  Section \ref{sec:discussion} concludes with a final discussion. Supplementary Materials provide the pseudo-code and additional details of the proposed algorithms.
 
\section{DATA DESCRIPTION~\label{sec:data_description}}

In this paper we focus on the process of blood cell formation described in the Introduction. We use the data generated by \cite{psaila2016}, as well as their classification of the $n=681$ megakaryocyte-erythroid progenitor (MEP) cells into three distinct sub-populations, identified via a principle component analysis. A close inspection of the expression patterns of the megakaryocyte-associated and erythroid-associated genes across the three sub-populations have led \cite{psaila2016} to the identification of the three groups as, respectively, the most undifferentiated state, which they name as ``precursor-MEP" (Pre-MEP, $n_1=255$ samples), and two committed lineages, which they name as ``erythroid-primed MEP'' (E-MEP, $n_2=241$) and ``megakaryocyte-primed'' MEP (MK-MEP, $n_3=185$). 

We analyse these data further, by elucidating common and specific features that characterize the most undifferentiated state and each of the two committed lineages at the level of cellular pathways among important genes. Of the 87 transcripts that are profiled by  \cite{psaila2016}, 11 transcripts are not expressed in at least one of the three groups and hence removed from this analysis, while 2 transcripts (\textit{B2M} and \textit{GAPDH}) are housekeeping genes and are used for data normalization. The remaining 74 genes are classified into two blocks: on one side the transcripts that code for membrane-bound receptors, their ligands, cytoplasmic signal transducers and metabolic enzymes (these will enter the model as $q=40$ external covariates), while, on the other side the transcripts, such as the cyclin-dependent kinases, that code for nuclear factors and master regulators of nuclear and cellular activities (resulting in a total of $p=34$ response variables). When a protein possesses a dual activity (e.g. it can be both membrane-bound and nuclear) the preference is given to the nuclear activity. Using the method described in this paper, we aim to recover the lineage-dependent regulatory networks, both at the level of the response variables, of the covariates and of the associations between the two, and to identify the key lineage-specific features among the many shared pathways. 

The expression of the transcripts in each cell is measured by RT-qPCR technology. As typical of this technology, data are only partially observed. The most common explanation for this in the context of RT-qPCR technology is that some samples fail to attain the minimum (pre-specified) signal intensity before reaching the maximum number of cycles  (denoted with Ct for threshold-cycle and with a maximum of 40 for the case of \cite{psaila2016}). 
In this case, the missing data are generated by a right-censoring mechanism, as their true Ct value is greater than the maximum limit of detection. This assumption underlies a number of methods for imputing missing data generated by RT-qPCR, starting from the simplest, and most common approach, of setting those values to the limit of detection, as in \citep{psaila2016}, to more sophisticated approaches that, for example, exploit the availability of technical or biological replicates \citep{boyer2013}, account for the uncertainty in the missing data mechanism \citep{sherina2020} or allow for the simultaneous imputation of all transcripts from their joint multivariate Gaussian distribution \citep{AugugliaroEtAl_BioStat_20}. 

\begin{figure}[t!]
\centering
\includegraphics[scale = 0.58]{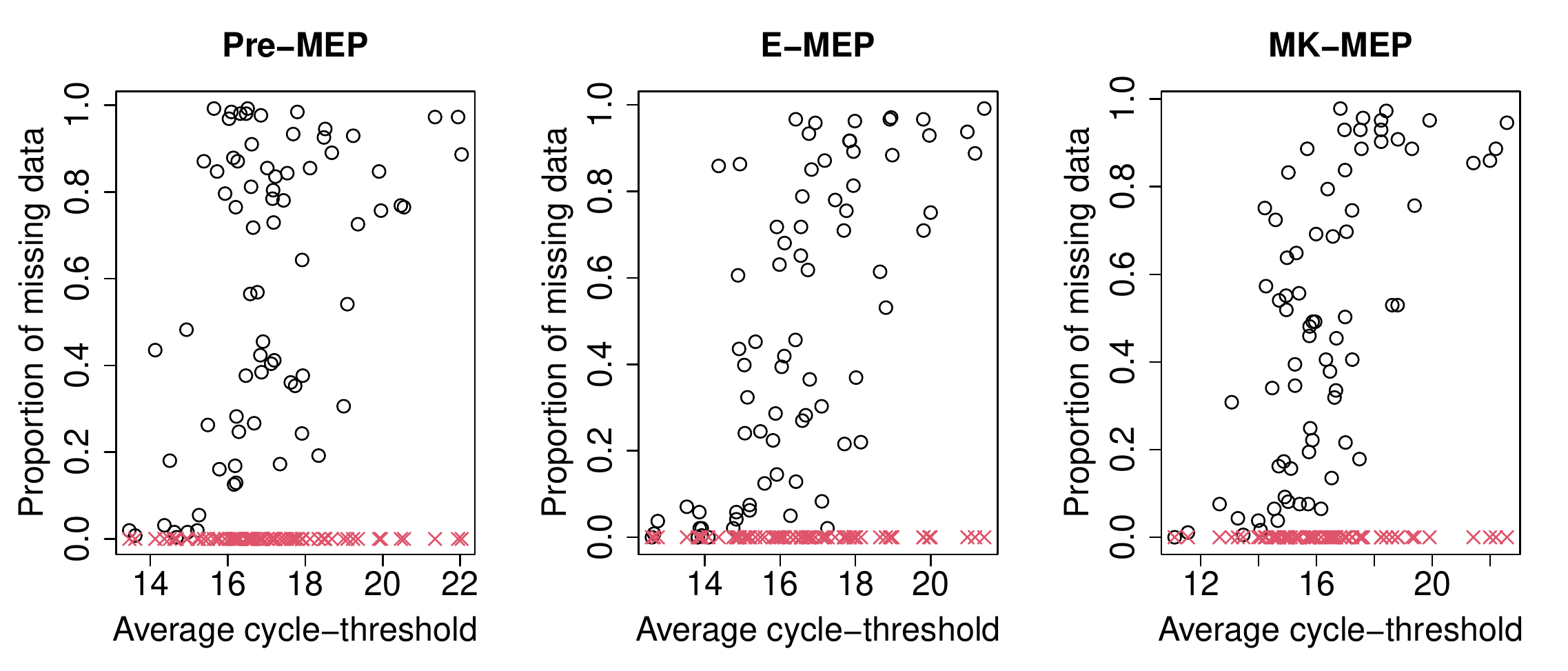}\\
\includegraphics[scale = 0.47]{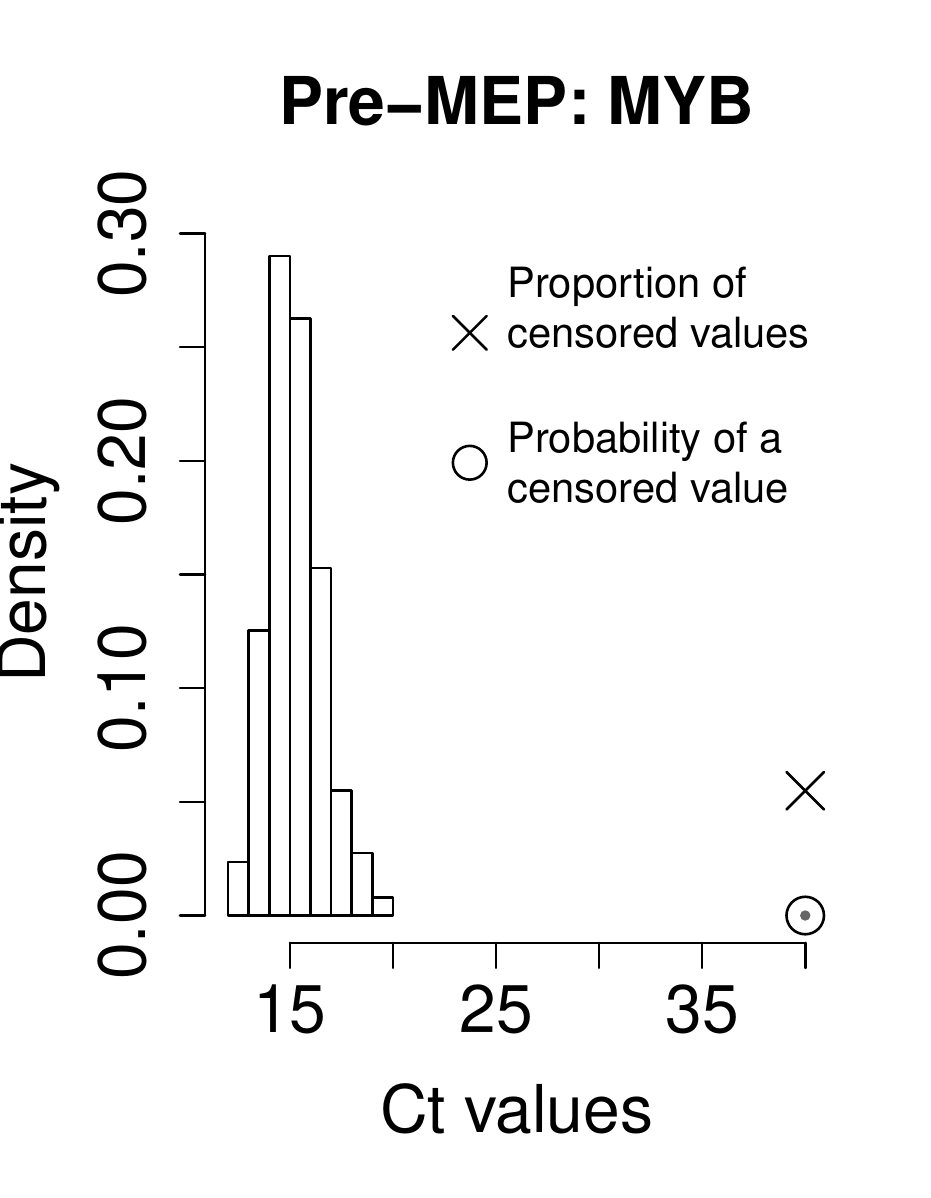}
\includegraphics[scale = 0.47]{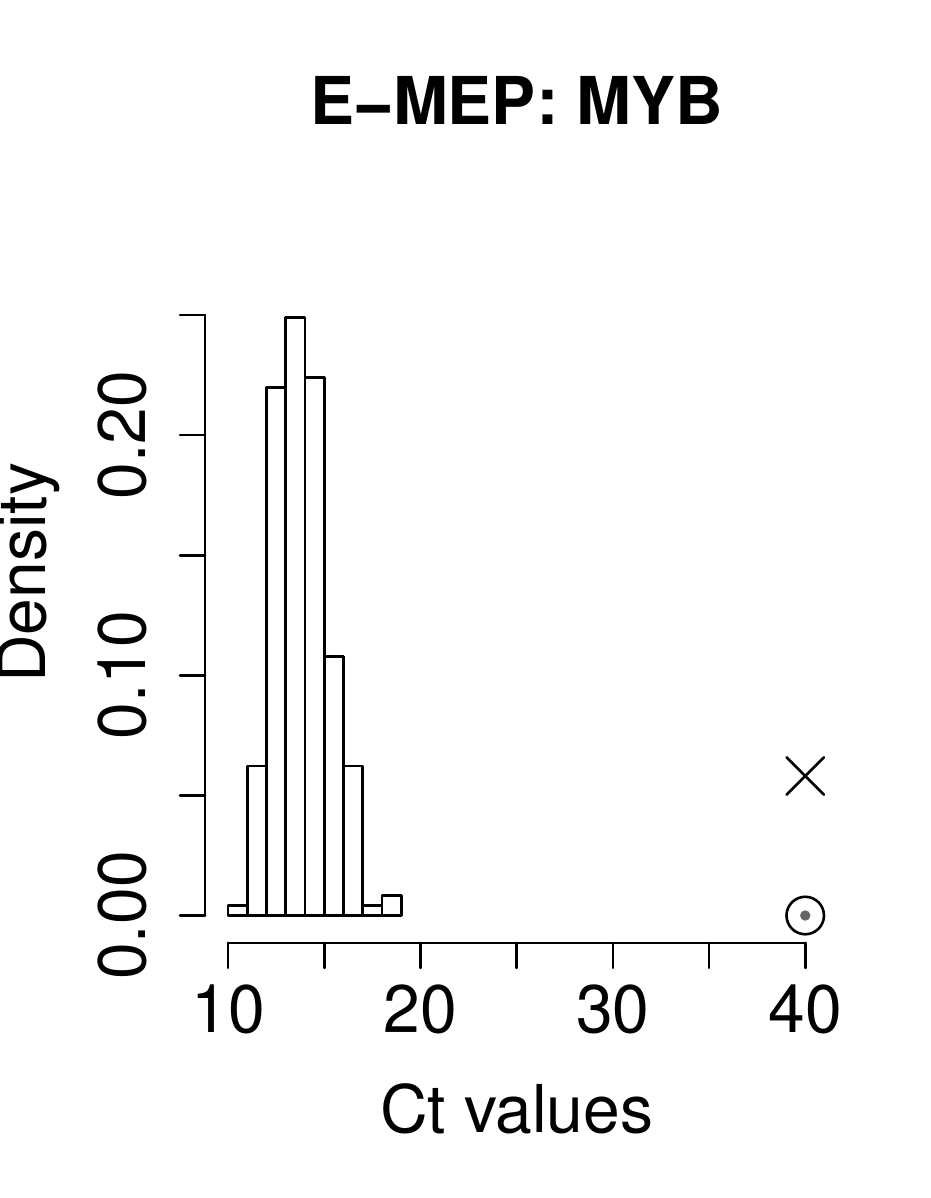}
\includegraphics[scale = 0.47]{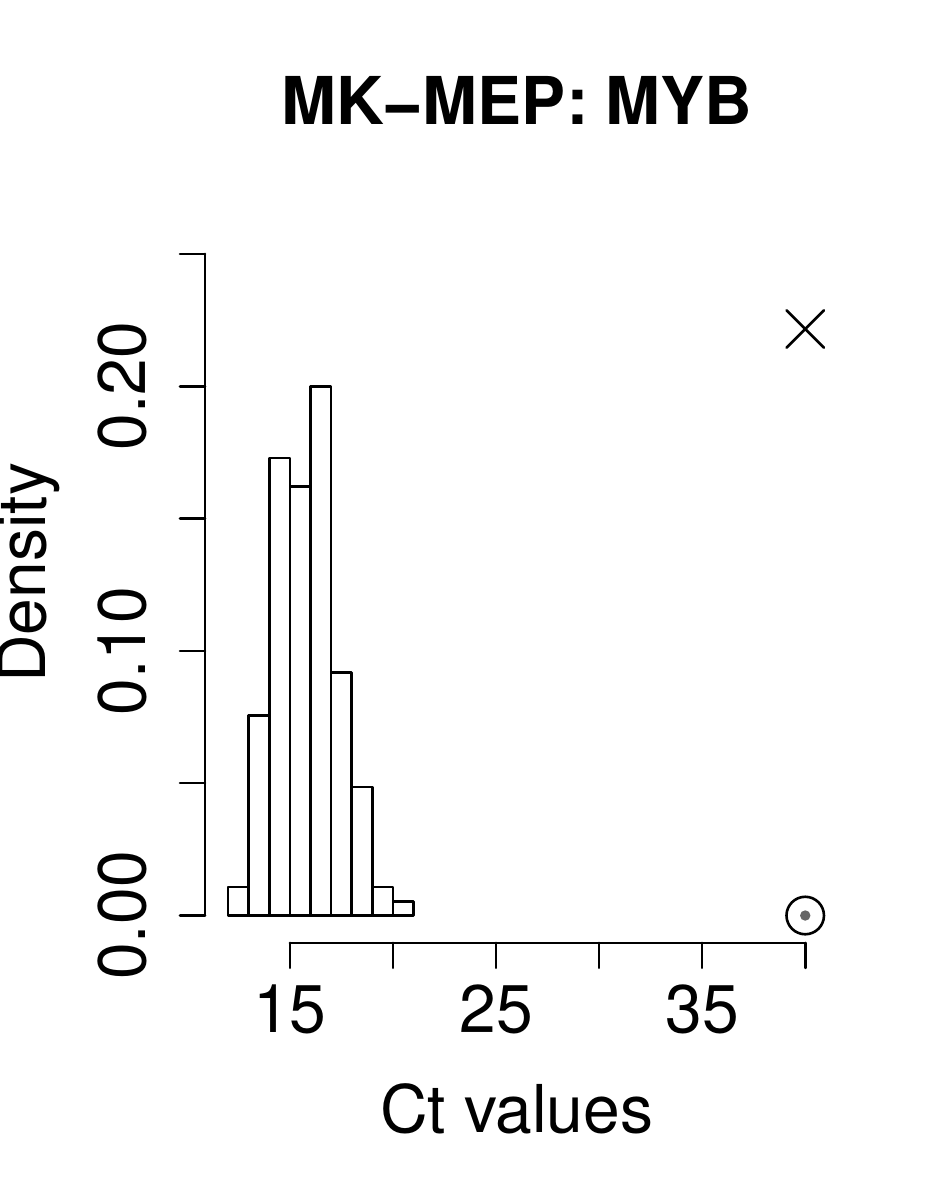}
\caption{Top: For each transcript, the average Ct value is plotted against the proportion of missing data (black circles) and the probability of observing censored values (red crosses), calculated by the marginal distribution fitted using the method described in this paper. Bottom: Histograms of the observed Ct values for a selected nuclear factor \textit{MYB} across the three sub-populations. Crosses indicate the proportion of censored values in the data, while circles indicate the expected probability of a censored value based on the fitted marginal model.\label{fig:ctvalues}}
\end{figure}

However, various authors have also pointed out how the censoring mechanism does not explain completely the extent of missingness typically observed in these data, with the percentage of non-detects far exceeding what one would expect from a censoring mechanism \citep{mccall2014}. This would suggest that some missing data may be the result of a failed amplification and therefore that their true value could in fact be also lower than the limit of detection. We find evidence of this in the data of \cite{psaila2016}. Indeed, Figure~\ref{fig:ctvalues} shows how, despite the percentage of missing data for each transcript tends to increase with the average cycle threshold value (black circles in the top panel), this percentage is far higher than what is implied by the associated marginal distribution (red crosses in the top panel). The bottom panel refers to one specific transcript and shows a typical example of how this discrepancy cannot be attributed solely to a potential miss-specification of the Gaussian distribution, that is assumed by our method, given the significant gap that is observed between the measured threshold values and the limit of detection. Taking all of this into consideration, we perform the analysis in this paper by assuming a missing-at-random data mechanism, for both the responses and the covariates. The method, however, will be described under both a missing-at-random and a censoring mechanism and can be applied in both cases.

\section{A GAUSSIAN GRAPHICAL MODEL FOR HETEROGENEOUS DATA} \label{sec:model}

In this section, we describe the GGM that will be used within each specific lineage. As mentioned in the Introduction, the model should capture the dependence structure within the nuclear factors, as well as that within the membrane-bound receptors and the associations between the two. This will motivate the selection of a specific form of parametrization for the model, which will result in a combination of a standard Gaussian graphical model at the level of membrane receptor activities and a conditional Gaussian graphical model at the level of the nuclear activities conditional on the membrane receptor activities.

Formally, let $\mathcal Y = (\mathcal Y_1,\ldots, \mathcal Y_p)^\top$ and $\mathcal X = (\mathcal X_1,\ldots, \mathcal X_q)^\top$ be $p$- and $q$-dimensional random vectors, respectively. In our setting, $\mathcal Y$ corresponds to the abundances of the $p$ nuclear activities and $\mathcal X$ corresponds to the abundances of the $q$ membrane receptor activities.  Let $\mathcal Z = (\mathcal X^\top, \mathcal Y^\top)^\top$ be the joint random vector, for which we assume a multivariate Gaussian distribution with mean and precision matrix given by, respectively:
\begin{equation}\label{eqn:block_par}
E(\mathcal Z) = \psi = %
\begin{pmatrix*}
\mu\\
\xi
\end{pmatrix*}, 
\quad\text{and}\quad%
V(\mathcal Z)^{-1} = \Psi = %
\begin{pmatrix*}[r]
\Omega + B\Theta B^\top& -B\Theta\\
-\Theta B^\top & \Theta
\end{pmatrix*}.
\end{equation}

The choice of this parametrization, which is different to the one proposed in earlier related work~\citep{AugugliaroEtAl_CompStat_20,SohnEtAl_PICAIS_12}, comes from the specific scientific questions of the current analysis. Indeed, the density function of $\mathcal Z$ given above can be factorized as follows:
\begin{equation}\label{eqn:LogLik_z}
\phi_z(z;\Upsilon) = \phi_{y\mid x}(y\mid x; \psi,B,\Theta)\,\phi_x(x;\mu,\Omega),
\end{equation}
where $\Upsilon = \{\psi, \Omega, B, \Theta\}$ is the full set of parameters and $\phi$ are multivariate Gaussian densities. In particular, $\phi_x$ describes the distribution of $\mathcal X$, essentially assuming a standard GGM for the $\mathcal X$ variables. Thus, the inverse of the covariance matrix of $\mathcal X$, here denoted with $\Omega = (\omega_{ij})$ and also called precision matrix, describes the regulatory network between the membrane-bound receptors. This is captured by a conditional independence graph. Indeed, using standard results about the multivariate Gaussian distribution, $\mathcal X_i$ and $\mathcal X_j$ are conditionally independent  given the remaining $\mathcal X$ variables, i.e. a missing link in the network, iff the corresponding precision value $\omega_{ij}$ is zero \citep{Lauritzen_book_96}. On the other hand, the density $\phi_{y\mid x}$ describes the distribution of $\mathcal Y$ conditional on $\mathcal X$, as in a conditional GGM, and is characterized by the probability density function:
\begin{equation}\label{eqn:mgd}
\phi(y\mid x; \psi,B, \Theta) = (2\pi)^{-\frac{p}{2}}|\Theta|^{\frac{1}{2}}\exp\left\{-\frac{1}{2} (y - \beta_0 - B^\top x)^\top\Theta(y -  \beta_0 - B^\top x) \right\},
\end{equation}
where $B = (\beta_1\cdots \beta_p)$ is the $q\times p$  regression coefficient matrix, describing the associations between the membrane receptor and nuclear activities, $\beta_0 = \xi - B^\top\mu$ is the p-dimensional vector of intercepts, while $\Theta = (\theta_{ij})$ is the inverse of the conditional covariance matrix of $\mathcal Y$ given $\mathcal X$. Also here, using standard results about the multivariate Gaussian distribution, it is possible to show that $\mathcal Y_i$ and $\mathcal Y_j$ are conditionally independent  given $\mathcal X$ and  all the remaining $\mathcal Y$ variables iff the corresponding precision value $\theta_{ij}$ is zero.

The parameters $\Upsilon = \{\psi, \Omega, B, \Theta\}$ are able to account for, as well as distinguish, the dependence structures coming from the different sources of data. In the next section we discuss sparse inference for the model proposed. Here we will also account for data heterogeneity in terms of differences in the parameters between the multiple cellular states.  
	
\section{SPARSE ESTIMATION UNDER MULTIPLE CONDITIONS AND MISSINGNESS} \label{sec:inference}
In this section, we develop sparse inference for the proposed model in the challenging setting where data are collected from different sub-populations, or generally experimental conditions, and where data are only partially observed either at the response or predictor levels. This is indeed the case of our applied setting, where there are three sub-populations of cells (Pre-MEP, E-MEP and MK-MEP) and missingness both on $\mathcal X$ and on $\mathcal Y$.

Suppose that $n$ observations are collected from $K\ge 2$  sub-populations. Let $z_{i,k} = (x^\top_{i,k}, y^\top_{i,k})^\top$ denote the $i$th realization of the random vector $\mathcal Z$ in the $k$th sub-population.  As discussed above,  $z_{i,k}$ may be only partially observed. Thus, it can be split into the observed and unobserved sub-vectors, which we denote with  $z_{i,k}^o$ and $z_{i,k}^{no}$, respectively. The way the elements of $z_{i}^{no}$ are treated depends closely on the probabilistic assumption used to model the missing data mechanism. The framework described in this paper assumes that the missing values come either from a censoring mechanism or from a missing-at-random mechanism. 

Each sub-population is modelled by a distinct but related GGM, defined as in~(\ref{eqn:block_par}). Let then $\Upsilon_k = \{\psi_k, \Omega_k, B_k, \Theta_k\}$ denote the set of parameters of the $k$th GGM. Under the assumption of independent sampling and following ~\cite{LittleEtAl_Book_02} in the handling of missing data, the average observed log-likelihood function is given by
\begin{equation}\label{eqn_ave_oll}
\bar\ell(\{\Upsilon\}) = %
\frac{1}{n}\sum_{k=1}^K\sum_{i = 1}^{n_k}\log\int_{D_{i,k}} \phi(z^o_{i,k}, z^{no}_{i,k}; \Upsilon_k) \mathrm{d}  z^{no}_{i,k},
\end{equation}
where $\{\Upsilon\} = \{\Upsilon_1,\ldots,\Upsilon_K\}$ is the full set of parameters and $n_k$ is the sample size of the $k$th sub-population. The region of integration for observation $i$ in condition $k$, denoted with $D_{i,k}$, is defined as the Cartesian product of sub-regions, say $D_{ij,k}$,  whose definition depends on whether $z_{ij,k}^{no}$ is missing-at-random or censored. In the first case, $D_{ij,k}$ is equal to $\mathds R$, while in the second case, denoting with $l_{j,k}$ and $u_{j,k}$ the lower and upper censoring values of variable $\mathcal Z_j$ in sub-population $k$ (typically fixed by the experimental setting used), $D_{ij,k}$ will be equal to $(-\infty, l_{j,k})$, if $z_{ij,k} = l_{j,k}$, or $(u_{j,k}, +\infty)$ if $z_{ij,k} = u_{j,k}$.

Maximum likelihood estimators, found by maximizing~(\ref{eqn_ave_oll}), are known to have a high variance when the sample sizes are not large enough. In addition, they do not allow to highlight specific forms of similarity among the estimated networks. To overcome the inferential problems just described and to gain a more informative description of the data generating process, in this paper we propose the following penalized estimator:
\begin{equation}\label{eqn:jcmglasso}
\{\widehat\Upsilon\} = \arg\max_{\{\Upsilon\}} \bar\ell(\{\Upsilon\}) - \lambda P_{\alpha_1}(\{B\}) - \rho \widetilde{P}_{\alpha_2}(\{\Theta\}) - \nu \widetilde{P}_{\alpha_3}(\{\Omega\}) ,
\end{equation}
where $\{B\} = \{B_1,\ldots, B_K\}$, $\{\Theta\} = \{\Theta_1\ldots, \Theta_K\}$ and $\{\Omega\} = \{\Omega_1,\ldots,\Omega_K\}$ are the three dependence structures of interest, across the $k$ sub-populations. We call this estimator the joint conditional glasso estimator (in short \texttt{jcglasso}), to distinguish it from the one of \cite{DanaherEtAl_JRSSB_14} which is typically referred to as joint glasso (in short \texttt{jglasso}). Aside from allowing the presence of missing data in the inferential procedure, our estimator includes also the influence of covariates. This creates an additional layer of complexity, as both the $B$ and $\Omega$ matrices vary across sub-populations, as with the precision matrix $\Theta$.

The rationale for choosing the penalty functions in (\ref{eqn:jcmglasso}) is to select convex functions that encourage sparsity in each matrix as well as specific forms of similarity across the regression coefficient matrices and the precision matrices. In particular, for the regression coefficient matrices, we propose the following sparse group lasso penalty function:
\begin{equation*}
P_{\alpha_1}(\{B\}) = \alpha_{1}\sum_{k=1}^{K}\sum_{h = 1}^p \|\beta_{h,k}\|_1 + (1 - \alpha_1)\sum_{h = 1}^p\left(\sum_{k=1}^{K} \|\beta_{h,k}\|_2^2\right)^{1/2},
\end{equation*}
where $\alpha_1 \in [0, 1]$ is a parameter controlling the trade-off between a weighted lasso and a weighted group lasso penalty function. Although the previous penalty function is computational appealing, as we will discuss in Section~\ref{sec:MStep}, in principle other possible convex penalty functions could be used, such as the fused lasso-type penalty. The first penalty function is aimed at identifying zero regression coefficients, i.e. selecting the important associations between $\mathcal X$ and $\mathcal Y$, whereas the second penalty function encourages a similar pattern of sparsity of the regression coefficients across the different sub-populations. As for the precision matrices, both those corresponding to the dependence structure of $\mathcal X$ and $\mathcal Y$, we follow the proposal of~\cite{DanaherEtAl_JRSSB_14}. In particular, $\widetilde{P}_{\alpha_2}(\{\Theta\})$, and similarly $\widetilde{P}_{\alpha_3}(\{\Omega\})$, can take one of the following two forms:
\begin{equation}\label{dfn:penthetafused-joint}
\widetilde{P}_{\alpha_2}(\{\Theta\}) = \alpha_2\sum_{k=1}^{K}\sum_{h\ne m}|\theta_{hm,k}| + (1 - \alpha_2)\sum_{k<k^\prime}\sum_{h,m}|\theta_{hm,k}-\theta_{hm,k^\prime}|,
\end{equation}
or
\begin{equation}\label{dfn:penthetagroup-joint}
\widetilde{P}_{\alpha_2}(\{\Theta\}) = \alpha_2 \sum_{k=1}^{K}\sum_{h\ne m}|\theta_{hm,k}| + (1 - \alpha_2)\sum_{h\ne m}\bigg( \sum_{k=1}^{K}\theta_{hm,k}^2\bigg)^{1/2},
\end{equation}
where $\alpha_2\in[0, 1]$. The fused lasso penalty~(\ref{dfn:penthetafused-joint}) encourages a stronger form of similarity between the precision matrices, encouraging some entries of $\widehat\Theta_1,\ldots,\widehat\Theta_K$ to be identical across the $K$ sub-populations. On the contrary, the group lasso penalty~(\ref{dfn:penthetagroup-joint}) encourages only a shared pattern of sparsity. 

Although the computation of the proposed estimator may look infeasible at first glance due to the large number of penalty functions, in the next section we will show how the parametrization~(\ref{eqn:block_par}) and the factorization~(\ref{eqn:LogLik_z}), leads to an extremely flexible and relatively easy to implement EM algorithm, where specific ADMM sub-routines can be run in parallel, enabling the study of high-dimensional datasets. Moreover, we will devise a computationally efficient strategy for model selection in the setting considered, where sparsity is imposed via a number of tuning parameters. 


\section{COMPUTATIONAL ASPECTS} \label{sec:algorithm}

The EM algorithm is grounded on the idea of repeating the expectation and maximization steps, until a convergence criterion is met. We discuss in details these two steps.

\subsection{The expectation step} 
The first step, also called E-Step, requires the calculation of the conditional expected value of the complete log-likelihood function using the current estimates of the parameters. As our model assumes that, for each $k = 1, \ldots, K$, the random vector $\mathcal Z_k$ follows a multivariate Gaussian distribution, which is a member of the exponential family, the E-Step reduces to the calculation of conditional expected values of the sufficient statistics~\citep{McLachlanEtAl_Book_08}. In our case, this results in imputed missing data (from the first moments) and imputed covariance matrices (from the second moments), which are then fed to the maximisation step.

In particular, firstly, the missing values in the $k$th dataset are replaced with the following conditional expected values:
\begin{equation}\label{eqn:Zhat_k}
\hat z_{ij, k} = E_{\widehat\Upsilon_k} (\mathcal Z_{ij, k}\mid z_{i,k}\in D_{i, k}), 
\end{equation}
where $\widehat\Upsilon_k$ denotes the current estimate of the parameters, and $ E_{\widehat\Upsilon_k} (\cdot \mid z_{i,k}\in D_{i, k})$ is the expected value operator computed with respect to the multivariate Gaussian distribution truncated over the region $D_{i, k}$. The resulting imputed dataset is denoted by $\widehat Z_k = (\widehat X_k\; \widehat Y_k)$. 

Secondly, the matrix of second moments
\begin{equation*}
\widehat C_k = n_k^{-1} \sum_{i = 1}^{n_k} E_{\widehat\Upsilon_k} (\mathcal Z_{i, k}\mathcal Z_{i, k}^\top\mid z_{i,k}\in D_{i, k}) = %
\begin{pmatrix*}[r]
\widehat C_{xx,k} & \widehat C_{xy,k}\\
\widehat C_{yx,k} & \widehat C_{yy,k}
\end{pmatrix*},
\end{equation*}
is computed. This requires the computation of the moments of a multivariate truncated Gaussian distribution, 
with complex numerical algorithms for the calculation of the integral of the multivariate normal density function (see \citet{GenzEtAl_JCGS_02} for a review) that soon become computationally infeasible for the problem considered. Alternative efficient solutions are to either use a Monte Carlo method or to replace the mixed moments by the products of the conditional expectations, essentially calculating only conditional means and conditional variances. This second approach, proposed by~\cite{GuoEtAl_JCGS_15}, was  used successfully in a number of papers, e.g. \citep{BehrouziEtAl_JRSSC_19,AugugliaroEtAl_BioStat_20} among others, and will be considered also for the present paper. 

The complete log-likelihood function resulting from the E-Step, following imputation, is called the $Q$-function and  represents the key element of the M-Step. In the following, this function will be denoted by $Q(\{\psi\}, \{\Omega\}, \{B\}, \{\Theta\})$ to emphasize its dependence on the different sets of parameters.

\subsection{The maximization step\label{sec:MStep}} 
The M-Step solves a new maximization problem obtained by replacing the marginal log-likelihood function in~(\ref{eqn:jcmglasso}) with the $Q$-function, leading to the penalised $Q$-function:
\begin{equation}\label{eqn:maxQ}
\max_{\{\Upsilon\}} Q(\{\psi\}, \{\Omega\}, \{B\}, \{\Theta\}) - \lambda P_{\alpha_1}(\{B\}) - \rho \widetilde{P}_{\alpha_2}(\{\Theta\}) - \nu \widetilde{P}_{\alpha_3}(\{\Omega\}).
\end{equation}
The particular structure of the $Q$-function for our problem allows to solve the maximization problem~(\ref{eqn:maxQ}) through the solution of the following sequence of three sub-maximization problems:
\begin{align}
& \max_{\{\psi\}}Q(\{\psi\}, \{\widehat\Omega\}, \{\widehat B\}, \{\widehat\Theta\})\label{eqn:maxQ_psi},\\
& \max_{\{\Omega\}}Q(\{\hat\psi\}, \{\Omega\}, \{\widehat B\}, \{\widehat\Theta\}) - \nu \widetilde{P}_{\alpha_3}(\{\Omega\})\label{eqn:maxQ_Omg},\\
& \max_{\{B\}, \{\Theta\}}Q(\{\hat\psi\}, \{\widehat\Omega\}, \{B\}, \{\Theta\}) - \lambda P_{\alpha_1}(\{B\}) - \rho \widetilde{P}_{\alpha_2}(\{\Theta\}).\label{eqn:maxQ_BTht}
\end{align}
Each of these sub-problems involves only a set of parameters at once, while other parameters are held to their current estimates. Of particular notice is the fact that the last two sub-problems are unrelated to each other, in the sense that they can be written as separate optimization sub-routines involving only one set of parameters and the current estimates. Thus, they can be solved in parallel, improving the overall efficiency of the proposed algorithm and making it applicable to large datasets.

We start our discussion by considering sub-problem~(\ref{eqn:maxQ_psi}). Based on the current estimates of the parameters and, using the parametrization of the model~(\ref{eqn:block_par}), the $Q$-function can be rewritten as:
\begin{equation*}
Q(\{\psi\}, \{\widehat\Omega\}, \{\widehat B\}, \{\widehat\Theta\}) = - \sum_{k = 1}^K f_k (\bar z_k - \psi_k)^\top\widehat\Psi_k(\bar z_k - \psi_k) + C,
\end{equation*}
where $\bar z_k = n_k^{-1} \widehat Z_k^\top 1_{n_k}$, $f_k = n_k/ (2n)$ and $C$ is a constant term. This then leads to the well-known optimal solution of sub-problem~(\ref{eqn:maxQ_psi}), given by:
\[
\hat\mu_k = n_k^{-1} \widehat X_k^\top 1_{n_k}\quad\text{and}\quad\hat\xi_k = n_k^{-1} \widehat Y_k^\top 1_{n_k},\qquad k = 1, \ldots, K.
\]

Given the new estimates $\{\hat\psi\}$, sub-problems~(\ref{eqn:maxQ_Omg}) and~(\ref{eqn:maxQ_BTht}) can be efficiently solved using the factorization~(\ref{eqn:LogLik_z}) of the joint density function of $\mathcal Z$ as a product of the conditional distribution of $\mathcal Y$ and the marginal distribution of $\mathcal X$. Indeed, thanks to this factorization, the $Q$-function admits the following additive structure:
\begin{multline}
Q(\{\hat\psi\}, \{\Omega\}, \{ B\}, \{\Theta\}) = Q_x(\{\hat\mu\}, \{\Omega\}) + Q_{y\mid x}(\{\hat\psi\}, \{ B\}, \{\Theta\}) = \\
= \sum_{k=1}^{K}f_k\left[\log\det\Omega_k - \mathrm{tr}\{\Omega_{k}\widehat S_{xx,k}\} \right] + \sum_{k=1}^{K}f_k\left[\log\det\Theta_{k} - \mathrm{tr}\{\Theta_{k}\widehat S_{y\mid x,k}(B_k)\}\right],\label{eqn:Qfun_v2}
\end{multline}
where
\begin{eqnarray*}
\widehat S_{xx,k} &=& \widehat C_{xx,k} - \hat\mu_k\hat\mu_k^\top,\\
\widehat S_{y\mid x,k}(B_k) &=& (\widehat C_{yy,k}  -\hat\xi_k\hat\xi_k^\top) - (\widehat C_{yx,k} - \hat\xi_{k}\hat\mu_k^\top)B_k - B_k^\top(\widehat C_{xy,k} - \hat\mu_k\hat\xi_{k}^\top)+ B_k^\top\widehat S_{xx,k} B_k\\
&=& \widehat S_{yy,k} - \widehat S_{yx,k} B_k - B_k^\top\widehat S_{xy,k} + B_k^\top\widehat S_{xx,k} B_k.
\end{eqnarray*}
The additivity of expression~(\ref{eqn:Qfun_v2}) justifies why problems~(\ref{eqn:maxQ_Omg}) and~(\ref{eqn:maxQ_BTht}) can be solved in parallel. Indeed, re-writing~(\ref{eqn:maxQ_Omg}) using~(\ref{eqn:Qfun_v2}) and only the terms dependent on $\{\Omega\}$, we have:
\begin{equation}\label{eqn:subprobJGL}
\{\widehat\Omega\} = \arg\max_{\{\Omega\}} \sum_{k=1}^{K}f_k\left[\log\det\Omega_k - \mathrm{tr}\{\Omega_{k}\widehat S_{xx,k}\} \right] - \nu \widetilde P_{\alpha_3}(\{\Omega\}), %
\end{equation}
which is equivalent to the definition of the \texttt{jglasso} estimator of~\cite{DanaherEtAl_JRSSB_14}. So we solve this optimization using the efficient ADMM algorithm proposed in that paper.

Finally, considering the sub-problem~(\ref{eqn:maxQ_BTht}) and using again~(\ref{eqn:Qfun_v2}), we have:
\begin{equation}\label{eqn:max_pQfun_ygivenx}
\max_{\{B\},\{\Theta\}}  \sum_{k=1}^{K}f_k\left[\log\det\Theta_{k} - \mathrm{tr}\{\Theta_{k}\widehat S_{y\mid x,k}(B_k)\}\right] - \lambda P_{\alpha_1}(\{B\}) - \rho \widetilde{P}_{\alpha_2}(\{\Theta\}),
\end{equation}
which represents an extension  of the conditional glasso problem studied by various authors~\citep{RothmanEtAl_JCGS_10,YinEtAl_AOAS_11,AugugliaroEtAl_CompStat_20} to the case of multiple conditions. Since for any fixed $\{\hat\psi\}$, the function $Q_{y\mid x}(\{\hat\psi\}, \{ B\}, \{\Theta\})$ is a bi-convex function of $\{B\}$ and $\{\Theta\}$, the maximization problem~(\ref{eqn:max_pQfun_ygivenx}) can be carried out by repeating the two sub-steps of estimation of $\{B\}$ and estimation of $\{\Theta\}$, respectively, until a convergence criterion is met. In particular, given the current estimate of $\{\widehat\Theta\}$ the set $\{B\}$ can be estimated  by solving the sub-problem:
\begin{equation}\label{eqn:subprob1}
\min_{\{B\}} \sum_{k = 1}^K f_k \mathrm{tr}\{\widehat{\Theta}_k\widehat{S}_{y\mid x,k}(B_k)\} + \lambda P_{\alpha_1}(\{B\}),
\end{equation}
whereas, given $\{\widehat{B}\}$, the set $\{\Theta\}$ is estimated by solving the sub-problem:
\begin{equation}\label{eqn:subprob2}
\max_{\{\Theta\}} \sum_{k=1}^{K}f_k\left[\log\det\Theta_{k} - \mathrm{tr}\{\Theta_{k}\widehat S_{y\mid x,k}(\widehat{B}_k)\}\right] - \rho \widetilde{P}_{\alpha_2}(\{\Theta\}).
\end{equation}
Similar to problem~(\ref{eqn:subprobJGL}), problems~(\ref{eqn:subprob1}) and (\ref{eqn:subprob2}) can be solved via ADMM algorithms. Indeed, problem~(\ref{eqn:subprob2}) corresponds to a \texttt{jglasso} where the standard empirical covariance for each condition is replaced by the conditional imputed covariance. On the other hand, problem~(\ref{eqn:subprob1}) requires further development and is discussed separately.

Going then into the detail of problem~(\ref{eqn:subprob1}), while it could be solved, in principle, by the multi-lasso algorithm of~\cite{AugugliaroEtAl_CompStat_20}, this algorithm tends to be unstable when the response vectors are highly correlated, as it is based on the idea of optimizing one column of each regression coefficient matrix at a time until a convergence criterion is met. This shortcoming motivated us to develop a dedicated ADMM. In particular, the scaled augmented Lagrangian for sub-problem~(\ref{eqn:subprob1}) is given by
\begin{equation*}
\begin{aligned}
L_\tau(\{B\},\{\Gamma\},\{U\}) = \sum_{k = 1}^K f_k \mathrm{tr}\{\widehat{\Theta}_k\widehat{S}_{y\mid x,k}(B_k)\} & + \lambda P_{\alpha_1}(\{\Gamma\}) +  
 \\
 &  +\frac{\tau}{2}\sum_{k = 1}^K ||B_k - \Gamma_k + U_k||_F^2 - \frac{\tau}{2}\sum_{k = 1}^K ||U_k||_F^2,
 \end{aligned}
\end{equation*}
where $U_k$ are dual matrices, $\tau > 0$ is a penalty parameter controlling the step size and $\|\cdot\|_F$ denotes the Frobenius norm. As described in~\cite{boyd2011distributed}, the ADMM algorithm is based on the idea of repeating the minimization of $L_\tau(\{B\},\{\Gamma\},\{U\})$ with respect to a set of parameters, while the remaining parameters are kept fixed to the previous values. Formally, the algorithm consists in repeating the following three steps:
\begin{itemize}
\item[] $\{B^{(i)}\} = \arg\min_{\{B\}} \sum_{k = 1}^K \left[ f_k \mathrm{tr}\{\widehat{\Theta}_k\widehat{S}_{y\mid x,k}(B_k)\} + \frac{\tau}{2}||B_k -\Gamma^{(i - 1)}_k +U^{(i - 1)}_k||_F^2\right],$
\item[] $\{\Gamma^{(i)}\} = \arg\min_{\{\Gamma\}}  \frac{\tau}{2}\sum_{k = 1}^K ||A_k - \Gamma_k||_F^2 + \lambda P_{\alpha_1}(\{\Gamma\})$,
\item[] $\{U^{(i)}\} = \{U^{(i - 1)}\} + \{B^{(i)}\} - \{\Gamma^{(i)}\}$,
\end{itemize}
until a convergence criterion is met. In the examples throughout the paper, we used $\tau = 2$ and declared convergence when $\sum_{k = 1}^K ||B_k^{(i)}-B_k^{(i-1)}||_1 / \sum_{k = 1}^K ||B_k^{(i-1)}||_1 < 10^{-5}$. Detailed derivations of the algorithm and pseudo-codes are reported in the Supplementary Materials.

\subsection{Tuning parameters selection}\label{sec:tuning_par}
The full inference is conducted for a selection of tuning parameters, namely $\alpha_1,\alpha_2,\alpha_3$, that allow to balance the two convex penalties, and $\lambda,\rho,\nu$, that control the degree of sparsity of the solution. For each combination of the six tuning parameters, the optimal model can be selected using the Bayesian Information Criterion (BIC), as suggested by many authors \citep{YinEtAl_AOAS_11,StadlerEtAl_StatComp_12,LiEtAl_JASA_12}. For the proposed model, this is defined by 
\begin{equation*}
\hbox{BIC} = -2n\bar\ell(\{\widehat\Upsilon\}) + \text{df} \sum_{k = 1}^K \log n_k,
\end{equation*}
where $\text{df}$ denotes the number of distinct non-zero estimated parameters in the regression matrices and precision matrices, and $\widehat{\Upsilon}$ is the MLE of the parameters constrained to the sparsity pattern of the solution at the given tuning parameters. Two problems appear: the first one is that the average log-likelihood in the BIC definition, defined in (\ref{eqn_ave_oll}), is not a direct output of the EM algorithm and its calculation adds a significant computational burden to the procedure; the second one is the large number of tuning parameters.

With regards to the first problem, using \cite{IbrahimEtAl_JASA_08} and \cite{AugugliaroEtAl_CompStat_20}, we replace the exact log-likelihood function with the $Q$-function used in the M-Step of the main algorithm, i.e. we use the following approximation:
\begin{eqnarray}\label{eqn:bic}
 \!\!\!\overline{\hbox{BIC}}\!\!&\!\!=\!\!&\!\!-2n Q(\{\widehat\Upsilon\}) + \text{df} \sum_{k = 1}^K \log n_k  \\
 \!\!\!&\!\!=\!\!&\!\!-2n Q_x(\{\hat\mu\},\{\widehat\Omega\})\!+\!\text{df}_x \sum_{k = 1}^K \log n_k\!-\!2n Q_{y\mid x}(\{\hat\psi\}, \{\widehat B\},\{\widehat\Theta\})\!+\!\text{df}_{y\mid x} \sum_{k = 1}^K \log n_k, \nonumber
\end{eqnarray}
where $\text{df}_x$ and $\text{df}_{y\mid x}$ correspond to the number of distinct non-zero parameters in $\{\widehat\Omega\}$ and in $(\{\widehat B\}, \{\widehat\Theta\})$, respectively.
Expression~(\ref{eqn:bic}) is computationally efficient as the two $Q$-functions are easily obtained as a byproduct of the EM algorithm.

With regards to the second problem, we have devised an efficient computational strategy on how to streamline the procedure. This is mainly based on two observations, which will be verified in the simulation section. The first one is that the mixing parameters $\alpha_1$, $\alpha_2$ and $\alpha_3$ have little impact on the selection of the optimal values of the main tuning parameters, $\lambda$, $\rho$ and $\nu$. Indeed, the simulation will show how different values of $\alpha$ cause a shift in goodness-of-fit measures across the other tuning parameters, but do not  change significantly the optimal points. One strategy is therefore to fix $\alpha_1$, $\alpha_2$ and $\alpha_3$ a priori to some realistic values, e.g. a high value if one expects a high sharing across conditions. The second observation is that Expression~(\ref{eqn:bic}) is conveniently split into two additive components, one concerning $\{\widehat\Omega\}$ and the other one regarding $\{\widehat B\}, \{\widehat\Theta\}$. The two components are not completely independent of each other when it comes to tuning parameter selection. Indeed, a selection of $\nu$ that minimizes the first component returns also a $\{\hat\mu\}$ that may affect the estimates of $\{\widehat B\}$ and $\{\widehat\Theta\}$ in the second component. However, since $\nu$'s main effect is to induce sparsity on $\Omega$, the simulation will show how $\{\hat\mu\}$ is mostly unaffected by the selection of $\nu$, with nearly constant values across $\nu$. From this, we conclude that $\{\widehat B\}$ and $\{\widehat\Theta\}$ are also mostly unaffected by the selection of $\nu$, which means that the two components of the BIC can be regarded as two separate sub-problems when it comes to tuning parameter selection, i.e. they can be minimized separately. 

While the first sub-problem, that of selecting $\nu$, is standard in the glasso literature, and can in fact be avoided  by setting $\nu=0$ if sparsity of $\Omega$ is not of interest and enough data are available, the second sub-problem requires the selection of two tuning parameters, $\lambda$ and $\rho$. As in the case of conditional graphical models, this requires the definition of a two dimensional grid, say $\bm{\lambda}=\{\lambda_{\min},\ldots,\lambda_{\max}\}\times\bm{\rho}=\{\rho_{\min},\ldots,\rho_{\max}\}$, over which to evaluate the goodness-of-fit of the model. When the sample size is large enough, one can let $\lambda_{\min}$ and/or $\rho_{\min}$ equal to zero, so that the maximum likelihood estimate is one of the points belonging to the coefficient path. On the other hand,  in a high-dimensional setting, one can use two values small enough to avoid the instability of the model. With respect to the largest values of the two tuning parameters, combining the results of Theorems 1 and 2 of  \cite{DanaherEtAl_JRSSB_14} and of Theorem 3 of \cite{AugugliaroEtAl_CompStat_20} the formulae for computing $\lambda_{\max}$ and $\rho_{\max}$ are given, respectively, by
\begin{align*}
\lambda_{\max} & = \bigg|\bigg|\bigg\{\sum_{k=1}^{K}\bigg(f_k\bm{\widehat S}_{xy,k}\bigg)_+^2\bigg\}^{1/2}\bigg|\bigg|_\infty,\\
\rho_{\max} & = 
\begin{cases}
\max\bigg\{\bigg|\bigg|f_1\bm{\widehat S}_{yy,1}\bigg|\bigg|^-_\infty,\ldots,\bigg|\bigg|f_K\bm{\widehat S}_{yy,K}\bigg|\bigg|^-_\infty\bigg\} & \mbox{for the fused lasso penalty~\ref{dfn:penthetafused-joint}}\\
\bigg|\bigg|\bigg\{\sum_{k=1}^{K}\bigg(f_k\bm{\widehat S}_{yy,k}\bigg)_+^2\bigg\}^{1/2}\bigg|\bigg|^-_\infty & \mbox{for the group lasso penalty~\ref{dfn:penthetagroup-joint}}
\end{cases}
\end{align*}
where $(\cdot)_+^2$  indicates the square of the maximum between zero and the argument and $\|\cdot\|^-_\infty$ indicates the maximum element in absolute value of the upper triangular part of its argument. Once $\lambda_{\max}$ and $\rho_{\max}$ are calculated, the entire coefficient path can be computed over a two dimensional grid $\bm{\lambda}\times\bm{\rho}$ using the estimate obtained for a given pair $(\lambda, \rho)$ as warm starts for fitting the next model. 

\section{SIMULATION STUDY~\label{sec:simulation}}
The global behavior of the proposed \texttt{jcglasso} estimator is studied in two different simulation studies. In the first one, we measure the performance of the method under several settings, selected with a view to evaluate the procedure that we have devised for tuning parameter selection. In the second one, the proposed method is compared with the joint graphical lasso~(\citealp{DanaherEtAl_JRSSB_14}) estimator, where censored data are imputed using their limit of detection and no covariates are included. 

For simulating data under different conditions, the following general setting is used. We set the number of groups ($K$) to 3, the sample size per condition ($n_k$) to $100$ and  a range of values for  the number of response variables ($p$), covariates ($q$), and missingness. Response and covariate data are simulated according to a multivariate Gaussian distribution, with  precision matrices having an underlying sparse graph structure, that is, $\theta_{h(h + j),k}$, with $h = 1, 6, 11, \ldots, p-4$ and $j = 1, \ldots,4$. The diagonal entries are fixed to $1$ while the non-zero precision values are sampled from a uniform distribution on the interval $[0.30, 0.50]$. The estimators are evaluated both in terms of their ability to recover the presence/absence of dependencies and in their accuracy at estimating the true precision and regression coefficient values. In the first case, we calculate the area under the precision-recall curve along a path of tuning parameters, while, in the second case, we evaluate the mean squared error for a specific choice of tuning parameters. We calculate these measures across $50$ simulated samples and as an average across the three conditions. Thus, we formally define the three measures by:
\begin{align*}
\hbox{Precision} &=  \frac{1}{K}\sum_{k=1}^{K}\frac{\hbox{number of $\hat\theta_{hm,k}\ne0$ and $\theta_{hm,k}\ne0$}}{\hbox{number of  $\hat{\theta}_{hm,k}\ne0$}}, \\
\hbox{Recall}&= \frac{1}{K}\sum_{k=1}^{K}\frac{\hbox{number of $\hat\theta_{hm,k}\ne0$ and $\theta_{hm,k} \ne 0$}}{\hbox{number of $\theta_{hm,k}\ne0$}},\\
\hbox{MSE} & = \frac{1}{K}\sum_{k=1}^{K}E \|\widehat\Theta_{k} - \Theta_{k}\|^2_F,
\end{align*} 
where $\widehat\Theta_k$ would be replaced by $\widehat\Omega_k$ or $\widehat B_k$ in the case of the network of covariates and matrix of regression coefficients, respectively.  Details specific to the individual simulations will be provided in the respective subsections.

\subsection{Evaluating the performance and tuning parameter selection of \texttt{jcglasso}}
In this first set of simulations, we evaluate the performance of the proposed method in a complex setting, characterized by high dimensionality in the response variables ($p=200$) with external covariates ($q=50$), and a high percentage of missingness. In particular, 40\% of the response variables are subject to right-censoring, with a censoring value set to $40$, and 40\% of the covariates are subject to missingness-at-random. For each of these variables, the probability of missing is set to 0.40, with the mean of the censored variables adjusted accordingly. The precision matrices across the three conditions, $\{\Theta\}$ and $\{\Omega\}$, are set as described before, while the three coefficient matrices $\{B\}$ are generated with a sparse structure, where only the first two rows are non-zero with coefficients sampled from a uniform distribution on the interval $[0.30,0.70]$.

For the simulation study, we consider two main scenarios. In the first scenario, we study the sensitivity of the selection of $\lambda$, $\rho$ and $\nu$ to the tuning parameters $\alpha_1,\alpha_2,\alpha_3$. In the second scenario, we study the sensitivity of  the selection of $\lambda$ and $\rho$ to the third tuning parameter $\nu$. Low sensitivity levels would support the choice of setting $\alpha$ values first (e.g. to some a priori realistic values), then selecting $\nu$ (and estimating $\{\Omega\}$) and, finally, selecting $\lambda$ and $\rho$ (and estimating $\{\Theta\}$ and $\{B\}$).

\begin{figure}[!t]
\centering
\makebox{
\includegraphics[scale=.75]{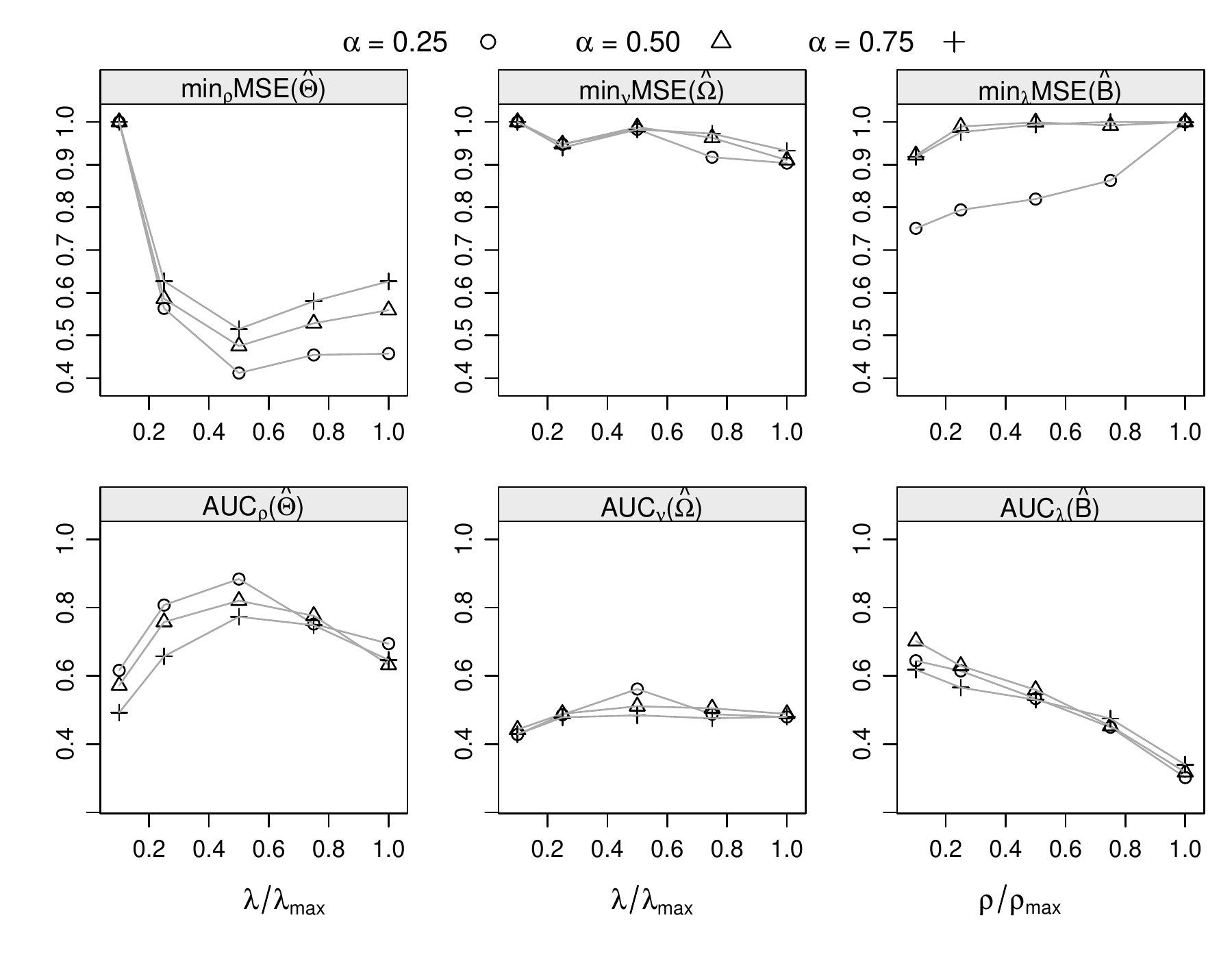}
}
\caption{Simulation study with $p=200$, $q=50$, $n_k=100$, 40\% of response variables with a 0.4 probability of censoring and 40\% of covariates with a 0.4 probability of missing-at-random. The three lines in each plot correspond to $\alpha_1=\alpha_2=\alpha_3\in\{0.25,0.50,0.75\}$. Upper and lower panels refer to the minimum MSE and the area under the precision-recall curves, respectively, corresponding to one tuning parameter ($\lambda$ or $\rho=\nu$) and varying the remaining tuning parameter along a sequence of values. To aid visualization, the MSE curves are scaled to a maximum value of one in each of the three settings.\label{fig:scenario1}}
\end{figure}

In the first scenario, we vary the $\alpha$ values according to three settings, namely $\alpha_1=\alpha_2=\alpha_3\in\{0.25,0.50,0.75\}$. We then explore the selection of the tuning parameters $\lambda$, $\rho$ and $\nu$ in these three cases. Here, for simplicity, we search for the optimal estimator in a grid of values where $\rho=\nu$, i.e. assuming the same sparsity in $\{\Omega\}$ and $\{\Theta\}$ for each condition, which is the case in this simulation. We explore a grid of values of the tuning parameters by first fixing the ratio $\lambda/\lambda_{\max}$ and then, for this fixed value, we compute the path across a decreasing sequence of $\rho$-values, namely $\rho/\rho_{\max}\in\{1.00, 0.75, 0.50, 0.25, 0.10\}$ setting $\nu=\rho$ along the sequence. Figure~\ref{fig:scenario1} summarizes the results in terms of parameter estimation (MSE) and network recovery (AUC of precision-recall curve). In particular, the top panel shows the MSE in the estimation of each matrix, calculated as the minimum along the corresponding path and across a sequence of values for the other tuning parameter. For example, the first plot reports the minimum MSE of $\{\widehat\Theta\}$ across the path of $\rho$ values for each fixed $\lambda$ value (and with $\nu=\rho$ along the sequence). Similarly, the bottom panel reports the AUC of the precision-recall curve of $\{\widehat\Theta\}$ across the path of $\rho$ values for each fixed $\lambda$ value (and with $\nu=\rho$ along the sequence). As observed in Section~\ref{sec:tuning_par}, the results show how the role of the tuning parameters $\alpha_1,\alpha_2,\alpha_3$ appears marginal in the selection of the remaining tuning parameters. Indeed, although the three curves in each plot may be different, the optimal point remains stable in all cases.

\begin{figure}[!t]
\centering
\makebox{\includegraphics[scale=.75]{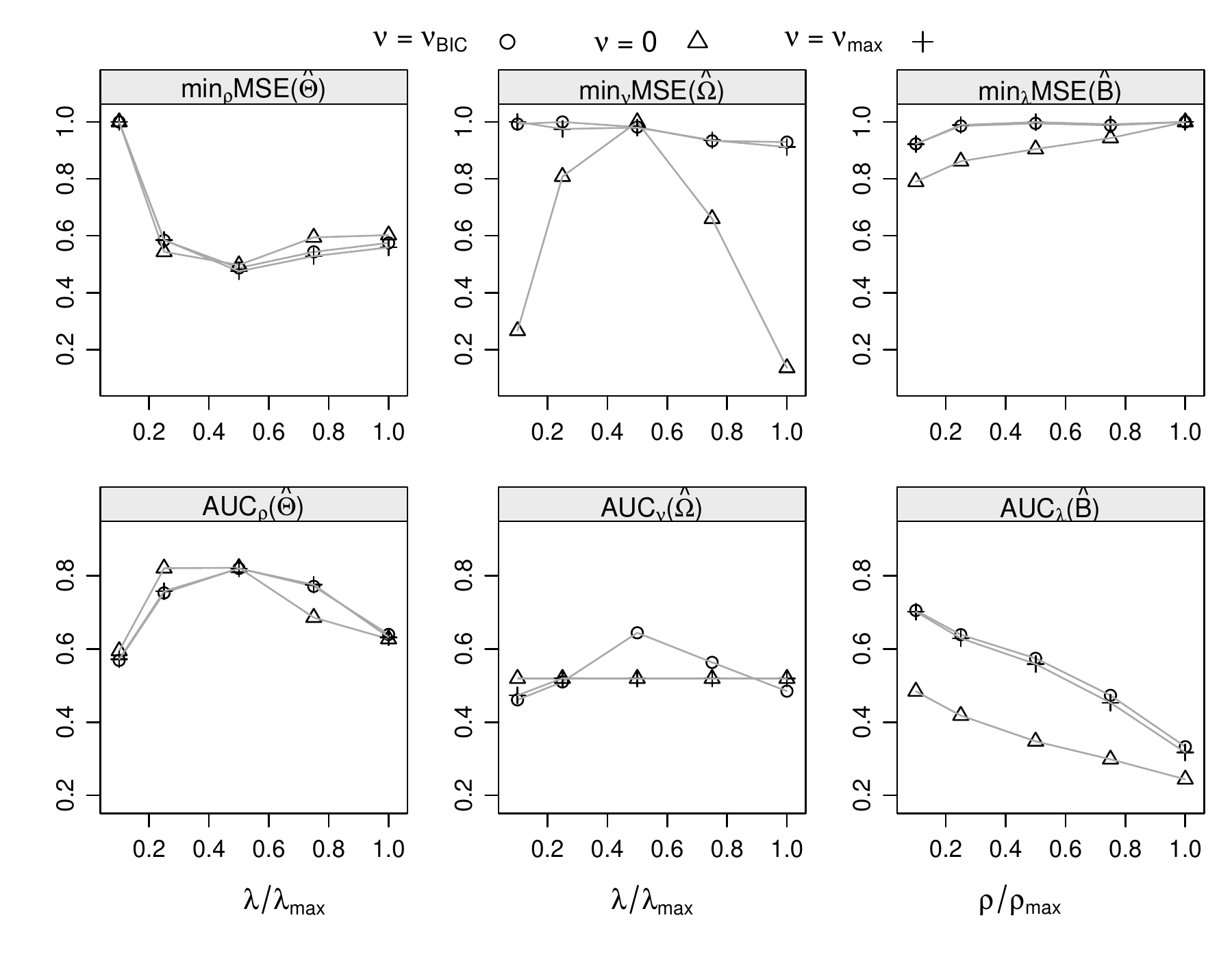}}
\caption{Simulation study with $p=200$, $q=50$, $n_k=100$, 40\% of response variables with a 0.4 probability of censoring and 40\% of covariates with a 0.4 probability of missing-at-random. We set $\alpha_1=\alpha_2=\alpha_3=0.5$. The three lines in each plot correspond to $\nu\in\{0,\nu_{\rm BIC},\nu_{\rm max}\}$, respectively. Upper and lower panels refer to the minimum MSE and the area under the precision-recall curves, respectively, corresponding to one tuning parameter ($\lambda$ or $\rho$) and varying the remaining tuning parameter along a sequence of values. To aid visualization, the MSE curves are scaled to a maximum value of one in each of the three settings.\label{fig:scenario2}}
\end{figure}

\begin{figure}[!htbp]
\centering
\makebox{\includegraphics[scale=.52]{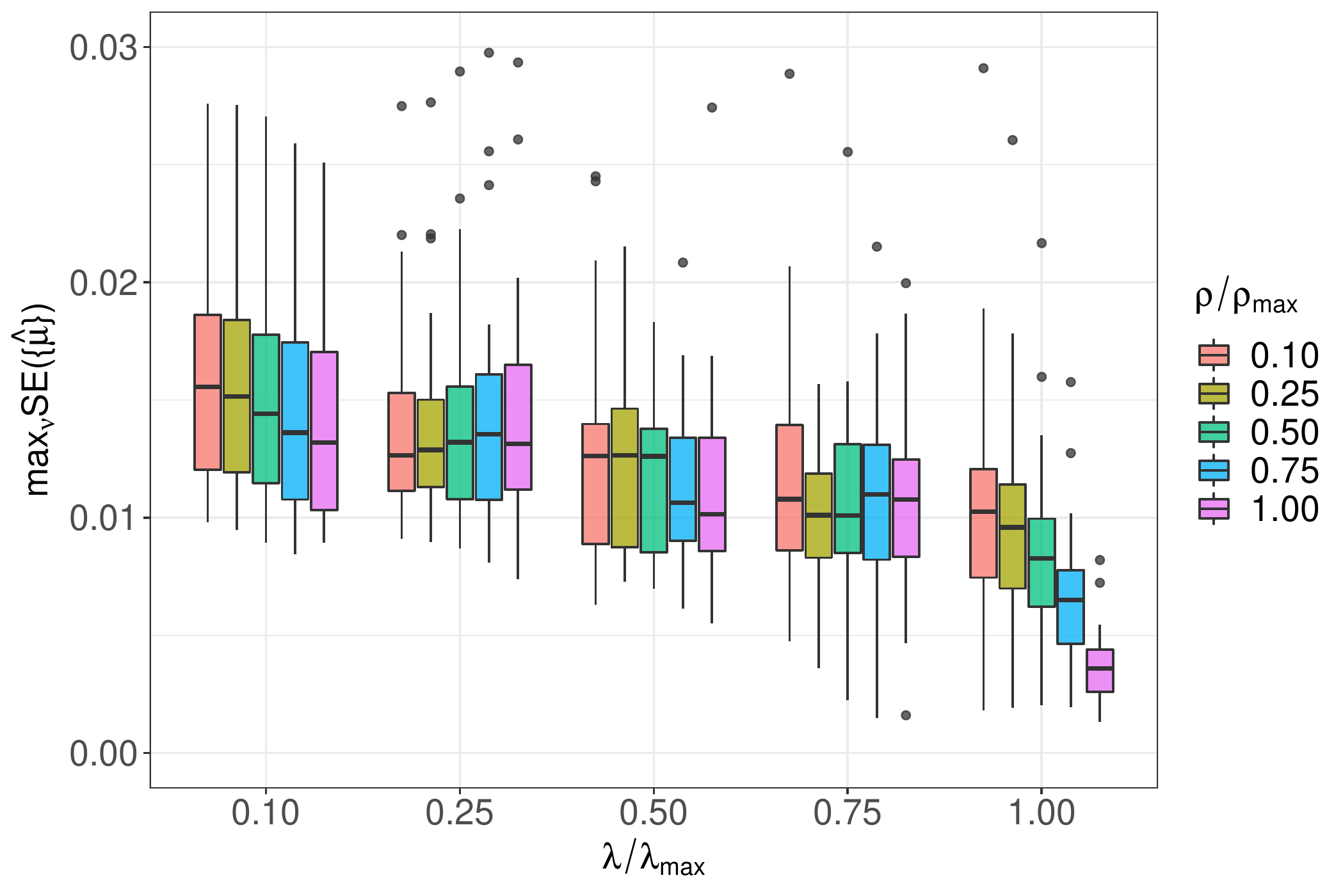}}
\caption{Simulation study with $p=200$, $q=50$, $n_k=100$, 40\% of response variables with a 0.4 probability of censoring and 40\% of covariates with a 0.4 probability of missing-at-random. We set $\alpha_1=\alpha_2=\alpha_3=0.5$ and show the boxplot of the maximum standard errors of $\{\widehat{\mu}_k\}$ across an equally spaced grid of $\nu$-values, by keeping fixed $\lambda/\lambda_{\max}$ and $\rho/\rho_{\max}$.\label{fig:scenario3}}
\end{figure}

In the second scenario, we set $\alpha_1=\alpha_2=\alpha_3=0.50$ and consider the sensitivity of the selection of $\lambda$ and $\rho$ to the selection of $\nu$. Figure~\ref{fig:scenario2} shows the results of this simulation. The same quantities as in Figure~\ref{fig:scenario1} are plotted also here. However, in this case, we fix $\nu$ to three levels (0, the BIC optimal value and the largest value) and study the selection of $\lambda$ and $\rho$ in these cases. The figure shows how the selection of the optimal $\lambda$ and $\rho$ values is largely unaffected by the value of $\nu$. The reason is that the expected values $\{\widehat{\mu}_k\}$ , that appears both in the estimation of $\{\Omega\}$ and in the estimation of $\{\Theta\}$ and $\{B\}$, are mostly unaffected by the value of $\nu$. 
Indeed, Figure~\ref{fig:scenario3} shows extremely low values for the maximum standard errors of  $\{\widehat{\mu}_k\}$ across an equally spaced grid of $\nu$-values, and across a range of fixed $\lambda/\lambda_{\max}$ and $\rho/\rho_{\max}$ values.

\subsection{Comparison with joint glasso}
In the last simulation, we compare our proposed estimator with \texttt{jglasso}~\citep{DanaherEtAl_JRSSB_14}, for which we use the implementation in the \texttt{R} package \texttt{JGL}. As this method does not account either for missing data or for external covariates, we simulate data without covariates and run \texttt{jglasso} on data where the censored data have been replaced with their limit of detection. 

Table~\ref{tab1} shows the performance of the methods across four different scenarios, ranging from a low dimensional case ($p=50, n_k=100$) with 20\% of the variables being censored ($M=10$) to a high dimensional case ($p=200, n_k=100$) with 40\% of variables censored ($M=80$). For the variables that are censored, the probability of censoring is fixed to 0.40. For each scenario, we simulate $50$ samples and in each simulation, we compute the coefficient path of \texttt{jcglasso} and \texttt{jglasso}, respectively, using the group lasso penalty in both cases. The path is computed using an equally spaced sequence of $\rho$-values and setting the $\alpha_2$ parameter equal to $0.50$. Table~\ref{tab1} reports averages of the performance measures across the three different conditions. The results show how \texttt{jcglasso} gives a better estimate of both the precision matrices (lower MSE at any given value of $\rho$) and the structure of the network (lower AUC values).
\begin{table}
\caption{Comparison between the proposed \texttt{jcglasso} and the joint graphical lasso (\texttt{jglasso}) under four different scenarios, with varying dimensions (p) and level of censoring (M). The first 5 columns refer to the mean squared error of the precision matrices $\{\Theta\}$, averaged across the 3 conditions and for five evenly spaced values of $\rho$. The last column refers to the average area under the precision-recall curves across the sequence of $\rho$-values. Standard errors are reported in brackets.\label{tab1}}
\centering
\fbox{
\begin{tabular}{lccccc p{.1cm} c}  
                    & \multicolumn{6}{c}{$\rho/\rho_\text{max}$ and $\alpha_2=0.50$}                    & \multirow{2}{*}{AUC} \\
\cmidrule{2-6}
                    & 0.10          & 0.25         & 0.50          & 0.75          & 1.00          & &                      \\
\midrule
\multicolumn{8}{l}{\textbf{Scenario 1: $p = 50$, $n_k = 100$, $M = 10$}} \\ 
\multicolumn{1}{r}{~\texttt{jcglasso}} & 3.22 (0.32) & 7.08 (1.25) & 11.91 (3.12) & 12.95 (3.84) & 12.90 (3.86) & & 0.99 (0.01)  \\   
\multicolumn{1}{r}{~\texttt{jglasso}} & 6.01 (0.66) & 10.08 (0.41) & 14.10 (1.81) & 14.52 (2.68) & 14.48 (2.73) & & 0.83 (0.02) \\[3mm] 
\multicolumn{8}{l}{\textbf{Scenario 2: $p = 50$, $n_k = 100$, $M = 20$}}\\
\multicolumn{1}{r}{~\texttt{jcglasso}} & 4.65 (0.54) & 8.06 (1.37) & 12.10 (3.18) & 12.92 (3.82) & 12.81 (3.83) & & 0.97 (0.01) \\   
\multicolumn{1}{r}{~\texttt{jglasso}} & 10.19 (1.46) & 13.67 (0.94) & 16.56 (0.93) & 16.84 (1.51) & 16.76 (1.57) & & 0.83 (0.02) \\[3mm]   
\multicolumn{8}{l}{\textbf{Scenario 3: $p = 200$, $n_k = 100$, $M = 40$}}\\
\multicolumn{1}{r}{~\texttt{jcglasso}} & 13.77 (1.18) & 31.79 (3.73) & 51.96 (10.44) & 54.97 (12.63) & 54.49 (12.58) & & 0.97 (0.01) \\ 
\multicolumn{1}{r}{~\texttt{jglasso}} & 24.96 (3.46) & 42.90 (1.48) & 59.97 (4.86) & 60.90 (7.78) & 60.80 (7.87) & & 0.88 (0.01) \\ [3mm]   
\multicolumn{8}{l}{\textbf{Scenario 4: $p = 200$, $n_k = 100$, $M = 80$}}\\
\multicolumn{1}{r}{~\texttt{jcglasso}} & 17.65 (2.37) & 34.85 (4.71) & 52.84 (10.72) & 55.18 (12.38) & 54.65 (12.26) & & 0.96 (0.01) \\ 
\multicolumn{1}{r}{~\texttt{jglasso}} & 45.32 (10.09) & 62.55 (9.16) & 75.53 (4.94) & 75.50 (2.65) & 75.32 (2.56) & & 0.81 (0.01)   \\  
\end{tabular}}
\end{table}

\section{INFERRING THE HUMAN HEMATOPOIETIC SYSTEM}\label{sec:realdata}

\begin{figure}[!t]
\centering
\makebox{
\includegraphics[scale = 0.37]{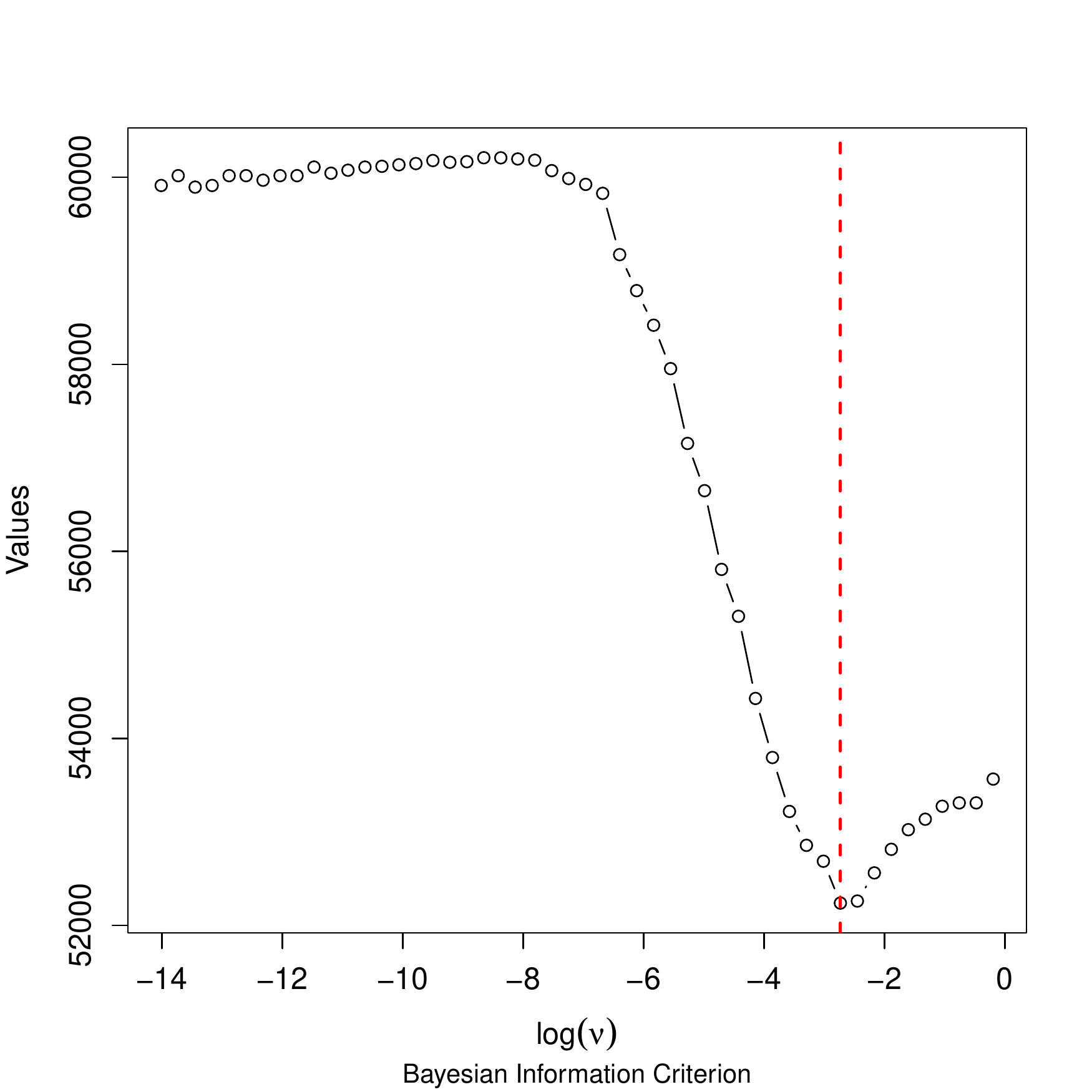}
\includegraphics[scale = 0.37]{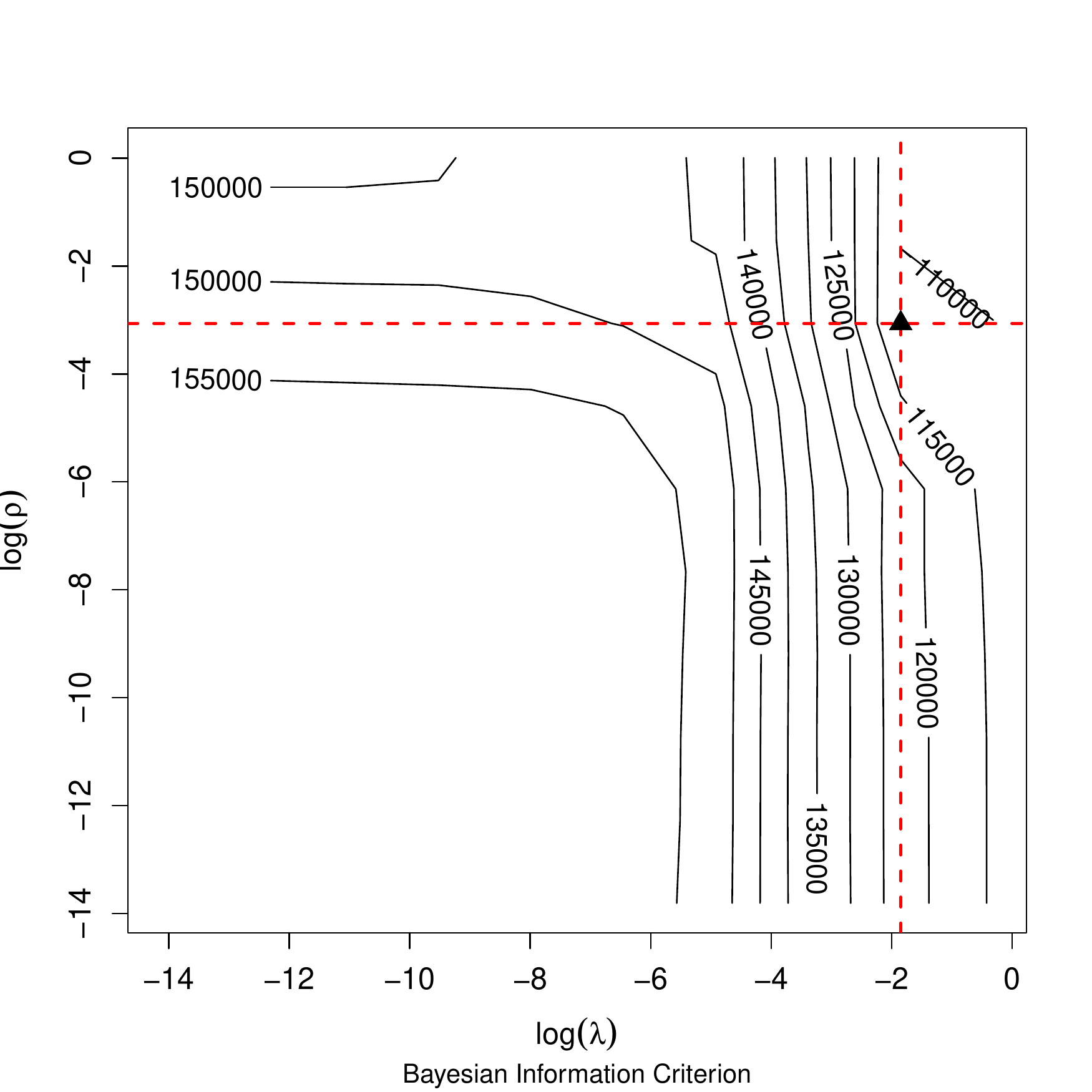}
}
\caption{Selection of the $(\nu, \lambda, \rho)$ tuning parameters on the real data analysis (after setting $\alpha_1=\alpha_2=\alpha_3=0.75$). Left: $\overline{\hbox{BIC}}$ across a path of 50 $\nu$ values; right: $\overline{\hbox{BIC}}$ contour lines on a two-dimensional grid of 100 $(\lambda,\rho)$ values given the $\{\widehat\Omega\}$ estimated at the optimal $\nu$. \label{fig:bic}}
\end{figure}

\begin{figure}[!t]
\centering
\makebox{\includegraphics[scale=.60]{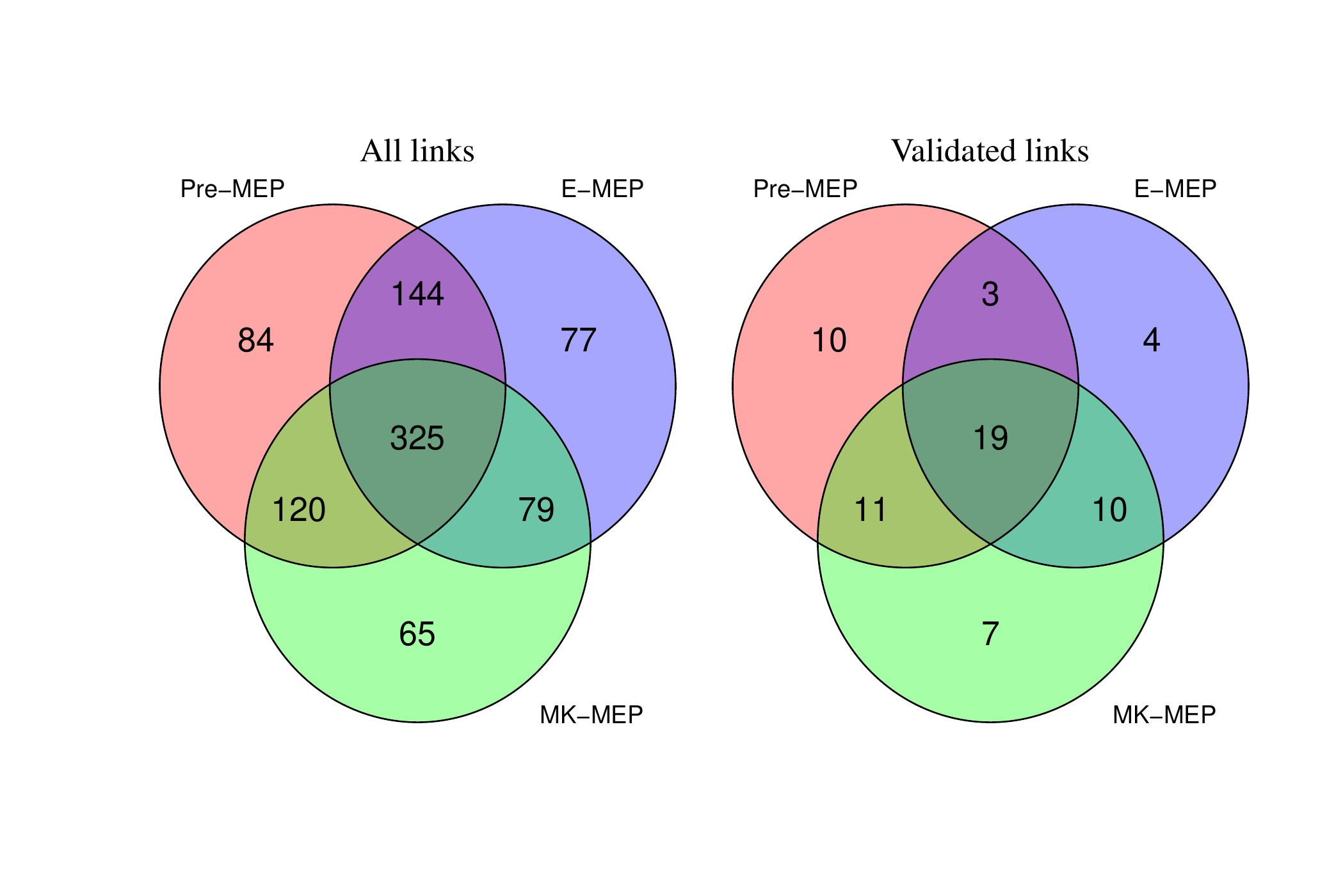}}
\caption{Venn diagrams reporting the number of edges common and specific to the three groups of megakaryocytes (left) and the number of those that have been experimentally validated (right). The latter is found using the IPA software.}
\label{fig:venn}
\end{figure}

\begin{figure}[!t]
\centering
\makebox{\includegraphics[scale=.60]{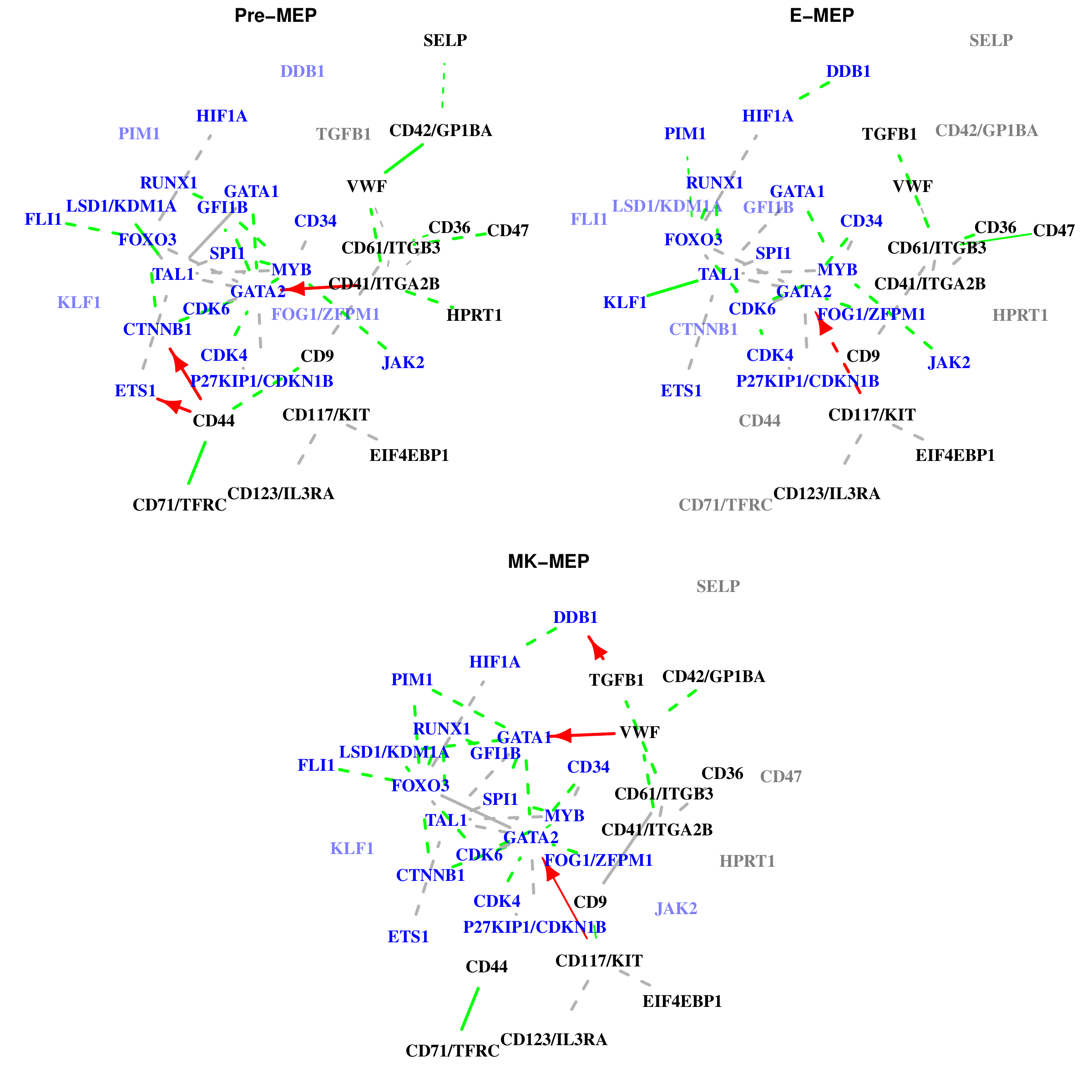}}
\caption{Regulatory networks of the nuclear and membrane receptors across the three subpopulations of cells. In each graph, blue, black and transparent nodes refer to membrane receptors, nuclear factors, and isolated nodes. Red arrows indicate edges from membrane receptors to nuclear factors; grey links indicate edges shared among the three megakaryocytes populations, while green links refer to specialised ones. Solid and dashed lines indicate positive and negative partial correlations, respectively, while the thickness of the line refers to the partial correlation value standardised to the maximum value inside the associated network.\label{fig:graphs}}
\end{figure}

The proposed \texttt{jcglasso} model is used to study the data generated by \cite{psaila2016} and gain insights on the process of blood cell formation described in the Introduction. Firstly, we address the problem of selecting the tuning parameters. We used the strategy identified and evaluated in the previous sections. In particular, we first set the three $\alpha$ parameters to $0.75$, since many common pathways are expected between the three sub-populations of cells and since the interest is in identifying the key differences between them. We then proceed by selecting the optimal $\nu$ value, i.e. inferring the sparsity structure of the 40 receptors for each condition ($\{\widehat\Omega\}$), followed by the selection of the pair $(\lambda,\rho)$, i.e.  the regulatory networks of the 34 nuclear activities ($\{\widehat\Theta\}$) and the associations between these and the membrane receptor activities ($\{\widehat B\}$). We use $\overline{\hbox{BIC}}$ for all tuning parameters, setting an equally spaced sequence of 50 values for the selection of $\nu$ and a $10\times10$ grid of values for the selection of $\lambda$ and $\rho$, given the optimal $\{\widehat\Omega\}$ identified before.  Figure~\ref{fig:bic} shows the $\overline{\hbox{BIC}}$ values across the path of solutions, and the optimal values that have resulted from this analysis ($\lambda=0.158$, $\rho=0.047$ and $\nu=0.065$). 
Looking now at the optimal networks, the number of unique non-zeros on the matrices associated to each population ($\widehat\Omega$, $\widehat\Theta$, $\widehat B$) is $\text{df}_\text{PRE}=781$ $(27.80\%)$, $\text{df}_\text{E}=733$ $(26.09\%)$ and $\text{df}_\text{MK}=697$ $(24.81\%)$, respectively (see Table~\ref{tabRes} for more details on the number of links at each level). Figure~\ref{fig:venn} (left) reports the number of edges that are common between, or specific within, each population. The right figure focusses only on those edges that have been experimentally validated. These are found using the ``connect'' tool of the Ingenuity Pathways Analysis software (Qiagen IPA, October 2021; \cite{kramer2014}). 
\begin{table}
\caption{Number of unique non-zeros (and density) of the optimal networks associated to each population ($\widehat\Omega$, $\widehat\Theta$, $\widehat B$). For each matrix, the first row refers to the overall number of links and the second focusses on those that have been experimentally validated.\label{tabRes}}
\centering
\fbox{
\begin{tabular}{cr cc p{.1cm} cc p{.1cm} cc}  
                    &  & \multicolumn{2}{c}{Pre-MEP} &  & \multicolumn{2}{c}{E-MEP} & & \multicolumn{2}{c}{MK-MEP} \\
\cmidrule{3-10}
                    &  & \#edges & Density           &  & \#edges & Density         & & \#edges & Density          \\
\midrule
$\widehat{\Theta}$ & All & ~94 & 16.76\% & & 113 & 20.14\% & & 104 & 18.54\% \\ 
               & Validated & ~26 & ~4.63\% & & ~26 & ~4.63\% & & ~33 & ~5.88\% \\[.1cm]
$\widehat{B}$      & All & ~18 & ~1.32\% & & ~26 & ~1.91\% & & ~29 & ~2.13\% \\ 
               & Validated & ~~3 & ~0.22\% & & ~~1 & ~0.07\% & & ~~3 & ~0.22\% \\[.1cm] 
$\widehat{\Omega}$ & All & 561 & 71.92\% & & 486 & 62.31\% & & 456 & 58.46\% \\ 
               & Validated & ~14 & ~1.79\% & & ~~9 & ~1.15\% & & ~11 & ~1.41\% \\[.1cm] 
\midrule
\textbf{Overall} & All & 673 & 24.92\% & & 625 & 23.14\% & & 589 & 21.81\% \\ 
                 & Validated & ~43 & ~1.57\% & & ~36 & ~1.31\% & & ~47 & ~1.72\% \\ 
\end{tabular}}
\end{table}
Figure~\ref{fig:graphs} focusses on these selected links, and distinguishes the links according to whether they refer to associations between the two levels (red arrows, corresponding to non-zeros in $B$) or to regulatory networks within the membrane receptors or nuclear factors (undirected edges, corresponding to non-zeros in $\Omega$ or $\Theta$). The latter are further distinguished between interactions that are common to at least two populations (in grey) versus those that are specific to one population (in green). A further characterization of an edge is in terms of a positive partial correlation (solid line) versus a negative partial correlation (dashed line).

From a biological point of view, the analysis highlighted a common shared network among all the differentiation states under investigation, composed by proteins that are crucial in the maintenance of hematopoietic stem cell properties. In particular, \textit{TAL1}, \textit{RUNX1}, \textit{GATA1} and \textit{GATA2} are hematopoietic master regulators \citep{vagapova2018}; while the interaction of \textit{MYB} with \textit{CDK4}, \textit{CDK6} and \textit{CDKN1B} is central for cell cycle regulation in hematopoietic stem cells \citep{matsumoto2013}.
Looking now at specific features in each sub-population, \textit{CD41} and \textit{CD44} stand out in the Pre-MEP network and their association to the most undifferentiated state is supported by literature \citep{shin2014}. Their connection with both \textit{CD71}/\textit{TFRC} (a marker of immature erythroid lineage) and \textit{VWF}, \textit{SELP}, \textit{CD61}, \textit{CD42} (megakaryocytic lineage markers - \cite{izzi2021}), is also coherent with the multipotent role of Pre-MEP cells. 
The more mature lineages (E-MEP and MK-MEP) share \textit{CD117}/\textit{KIT}. Indeed, low c-Kit expression has been associated with more immature forms \citep{shin2014}. Noteworthy, \textit{CD117}/\textit{KIT} is positively correlated with the \textit{GATA2} expression in MK-MEP, while it is negatively correlated in E-MEP. 
A unique feature of the E-MEP is the presence of \textit{KLF1}, which is indeed known to be necessary for the proper maturation of erythroid cells \citep{brown2002}. Instead, \textit{VWF} and \textit{TGFB1} are specific to the MK-MEP network. Indeed, \textit{TGFB1} inhibits the differentiation towards the erythrocytic lineage in favor of the megakaryocytic one \cite{blank2015}. 
Overall, it seems that \textit{CD117}/\textit{KIT} triggers the differentiation towards the default E-MEP lineage \citep{psaila2016}, while the commitment towards the MK-MEP lineage requires further support \citep{izzi2021}.

\section{CONCLUSION} \label{sec:discussion}
Motivated by the study of blood cell formation, and the search for regulatory mechanisms that underlie the differentiation of progenitor stem cells into more mature blood cells, we have proposed a complex graphical model that allows to infer interactions both within and between proteins belonging to different types, here nuclear factors and membrane receptors that are known to be involved in human hematopoiesis, and specific to each cellular lineage.  We have devised a computationally efficient strategy for the inference of the network of dependencies in a highly non-standard setting, characterized by high dimensionality, both at the level of responses and covariates, the presence of distinct sub-populations of cells, and missingness, which is typical of RT-qPCR data. The algorithm is based on a careful combination of alternating direction method of multipliers algorithms, that allow to achieve sparsity in the  networks within and the associations between the membrane-bound receptors and the nuclear factors specific to each lineage as well as a high sharing between networks across the different sub-populations. The approach 
can be used on similar applications where network inference is conducted from high-dimensional, heterogeneous and partially observed data.

\section*{ACKNOWLEDGEMENTS}

Luigi Augugliaro and Gianluca Sottile gratefully acknowledge financial support from the University of Palermo (FFR2021). Gianluca Sottile acknowledges support by the Italian Ministry of University and Research (MUR) through the project PON-AIM “Attraction and International Mobility”: AIM1873193-2 activity 1. Claudia Coronnello and Walter Arancio acknowledge support by Regione Siciliana, through the PO FESR action 1.1.5, project OBIND N.086202000366—CUP G29J18000700007.

\section*{CODE AVAILABILITY STATEMENT}

The computational approach presented in the paper will be implemented in the {\texttt R} package {\texttt cglasso}. The code for replicating the analysis presented in this paper is openly available at the following \texttt{GitHub} repository \url{https://github.com/gianluca-sottile/Hematopoiesis-network-inference-from-RT-qPCR-data}.

\bibliographystyle{rss}	
\bibliography{cglasso-jrssc}

\end{document}


\section{MULTI-LASSO ADMM ALGORITHM\label{apx:multilasso_admm}}

As discussed in Section~5.2, we solve sub-problem~(5.16) by a dedicated ADMM algorithm~\citep{boyd2011distributed}. We start by rewriting the original problem as the following constrained problem:
\begin{equation*}
\begin{aligned}
\min_{\{B\},\{\Gamma\}} & \sum_{k = 1}^K f_k \mathrm{tr}\{\widehat{\Theta}_k\widehat{S}_{y\mid x,k}(B_k)\} + \lambda P_{\alpha_1}(\{\Gamma\}),\\
\hbox{s.t.} \quad & B_k - \Gamma_k = 0, \qquad k = 1,\ldots,K,
\end{aligned}
\end{equation*}
whose corresponding scaled augmented Lagrangian takes the form:
\begin{equation*}
\begin{aligned}
L_\tau(\{B\},\{\Gamma\},\{U\}) = \sum_{k = 1}^K f_k \mathrm{tr}\{\widehat{\Theta}_k\widehat{S}_{y\mid x,k}(B_k)\} & + \lambda P_{\alpha_1}(\{\Gamma\}) +  
 \\
 &  +\frac{\tau}{2}\sum_{k = 1}^K ||B_k - \Gamma_k + U_k||_F^2 - \frac{\tau}{2}\sum_{k = 1}^K ||U_k||_F^2,
 \end{aligned}
\end{equation*}
where $U_k$ are dual matrices, $\tau > 0$ is a penalty parameter and $\|\cdot\|_F$ denotes the Frobenius norm. The ADMM algorithm consists in repeating the following three steps:
\begin{itemize}
\item[(a)] $\{B^{(i)}\} = \arg\min_{\{B\}} L_\tau(\{B\},\{\Gamma^{(i-1)} \},\{U^{(i-1)}\})$
\item[(b)] $\{\Gamma^{(i)}\} = \arg\min_{\{\Gamma\}} L_\tau(\{B^{(i)}\},\{\Gamma\},\{U^{(i-1)}\})$
\item[(c)] $\{U^{(i)}\} = \{U^{(i - 1)}\} + \{B^{(i)}\} - \{\Gamma^{(i)}\}$
\end{itemize}
until a convergence criterion is met. We now discuss the first two steps in detail. Rewriting the minimization problem in Step~(a) as follows:
\[
\min_{\{B\}}  \sum_{k = 1}^K \left[ f_k \mathrm{tr}\{\widehat{\Theta}_k\widehat{S}_{y\mid x,k}(B_k)\} + \frac{\tau}{2}||B_k -\Gamma^{(i - 1)}_k +U^{(i - 1)}_k||_F^2\right],
\]
we see that $\{B^{(i)}\}$ is the solution of $K$ distinct minimization problems whose solutions can be computed in closed form. In order to see that, consider the system of estimating equations defined by letting the derivative of the $k$th objective function equal to zero:
\begin{equation*}
\begin{aligned}
\partial_{B_k}  [ f_k \mathrm{tr} \{ \widehat{\Theta}_k\widehat{S}_{y\mid x,k}(B_k) & \}  +  \frac{\tau}{2}  ||B_k - \Gamma^{(i - 1)}_k +U^{(i - 1)}_k||_F^2] = \\
& = -2f_k\widehat{S}_{xy,k}\widehat{\Theta}_k + 2f_k\widehat{S}_{xx,k}B_k\widehat{\Theta}_k+\tau(B_k-\Gamma_k^{(i-1)} +U_k^{(i-1)})  = 0.
\end{aligned}
\end{equation*}
Using the $\mathrm{vec}(\cdot)$ and the well-known properties of the Kronecker product, denoted by $\otimes$, the previous system can be rewritten  as:
\begin{equation*}
\left\{ 2f_k(\widehat{\Theta}_k\otimes\widehat{S}_{xx,k}) + \tau\mathrm{I}\right\}\mathrm{vec}(B_k) - \mathrm{vec}\left(2f_k\widehat{S}_{xy,k}\widehat{\Theta}_k + \tau(\Gamma_k^{(i-1)} - U_k^{(i-1)})\right) = 0,
\end{equation*}
from which, we have:
\[
\mathrm{vec} (B^{(i)}_k) = \left\{ 2f_k(\widehat{\Theta}_k\otimes\widehat{S}_{xx,k}) + \tau I \right\}^{-1} \mathrm{vec}\left(2f_k\widehat{S}_{xy,k}\widehat{\Theta}_k + \tau(\Gamma_k^{(i-1)} - U_k^{(i-1)})\right).
\]
With regards to Step~(b), rewriting the corresponding minimization problem as follows:
\[
\min_{\{\Gamma\}} \frac{\tau}{2}\sum_{k = 1}^K ||A_k - \Gamma_k||_F^2 + \lambda P_{\alpha_1}(\{\Gamma\}),
\]
where $A_k = B^{(i)}_k + U^{(i - 1)}_k$, one can see that the minimization problem is formally equivalent to the problem (10) in~\cite{DanaherEtAl_JRSSB_14}. Thus it admits the following closed form solution:
\[
\Gamma_k^{(i)} = S(A_k, \alpha_1\lambda/\tau)\left[1-\frac{(1-\alpha_1)\lambda/\tau}{\{\sum_{k = 1}^K S(A_k, \alpha_1\lambda/\tau)^2\}^{1/2}}\right]_+,
\]
where the expression is intended elementwise and $S(\cdot,\cdot)$ denotes the soft-thresholding operator. Algorithm~\ref{algo:admm_lag} reports the pseudo-code of the proposed multi-lasso ADMM algorithm. Algorithm~\ref{algo:EM_cond_cglasso_type1} concludes this section by reporting the pseudo-code of the proposed EM algorithm summarising all its sub-routines.

\begin{algorithm}[!htbp]
\caption{Pseudo-code of the proposed multi-lasso ADMM algorithm\label{algo:admm_lag}}
\begin{algorithmic}[1]
\State Let $\{B\},\{\Gamma\},\{U\}$ be the initial estimates and select a scalar $\tau\ge0$
\State $i = 0$
\Repeat
\State $i = i + 1$
\For{$k=1\ldots K$}
\State $\mathrm{vec} (B^{(i)}_k) = \left\{ 2f_k(\widehat{\Theta}_k\otimes\widehat{S}_{xx,k}) + \tau I \right\}^{-1} \mathrm{vec}\left(2f_k\widehat{S}_{xy,k}\widehat{\Theta}_k + \tau(\Gamma_k^{(i-1)} - U_k^{(i-1)})\right)$
\EndFor
\For{$k=1\ldots K$}
\State $\Gamma_k^{(i)} = S(B^{(i)}_k + U^{(i - 1)}_k, \alpha_1\lambda/\tau)\left[1-\frac{(1-\alpha_1)\lambda/\tau}{\{\sum_{k = 1}^K S(B^{(i)}_k + U^{(i - 1)}_k, \alpha_1\lambda/\tau)^2\}^{1/2}}\right]_+$
\State $U_k^{(i)} = U_k^{(i-1)}+B_k^{(i)}-\Gamma_k^{(i)}$ 
\EndFor
\Until{a convergence criterion is met}
\end{algorithmic}
\end{algorithm}

\begin{algorithm}[!htbp]
\caption{Pseudo-code of the proposed EM algorithm\label{algo:EM_cond_cglasso_type1}}
\begin{algorithmic}[1]
\State Let $\{\widehat\Upsilon\}$ be the initial estimates
\Repeat
\For{$k=1\ldots K$}
\State use ${\widehat\Upsilon_k}$ and identities~(3.1) to compute $\widehat{\psi}_{k}$ and $\widehat\Psi_{k}$
\State use $\widehat{\psi}_{k}$ and $\widehat\Psi_{k}$ to update $\widehat Z_k$ and $\widehat C_k$
\EndFor
\State $\{\hat\psi\} =\arg\min_{\{\psi\}}  \sum_{k = 1}^K f_k (\bar z_k - \psi_k)^\top\widehat\Psi_k(\bar z_k - \psi_k)$
\State $\{\widehat\Omega\} = \arg\max_{\{\Omega\}} \sum_{k=1}^{K}f_k\left[\log\det\Omega_k - \mathrm{tr}\{\Omega_{k}\widehat S_{xx,k}\}\right] - \nu \widetilde{P}_{\alpha_3}(\{\Omega\})$
\Repeat
\State $\{B\} = \arg\min_{\{B\}} \sum_{k=1}^{K} f_k \mathrm{tr}\{\widehat\Theta_k S_{y\mid x,k}(B_k)\} + \lambda P_{\alpha_1}(\{B\})$
\State $\{\widehat\Theta\} = \arg\max_{\{\Theta\}} \sum_{k=1}^{K}f_k\left[\log\det\Theta_{k} - \mathrm{tr}\{\Theta_{k}S_{y\mid x,k}(\widehat{B}_k)\}\right] - \rho \widetilde{P}_{\alpha_2}(\{\Theta\})$
\Until{convergence criterion is met}
\Until{convergence criterion is met}
\end{algorithmic}
\end{algorithm}

\bibliographystyle{rss}	
\bibliography{cglasso-jrssc}